\documentclass{article}
\usepackage{amssymb}
\usepackage{graphicx}
\usepackage{color}
\usepackage[english]{babel}

\usepackage{cite}

\usepackage{epsf}
\newcommand{\be}{\begin{equation}}
\newcommand{\ee}{\end{equation}}
\newcommand{\bea}{\begin{eqnarray}}
\newcommand{\eea}{\end{eqnarray}}

\newcommand{\eeee}{\includegraphics[width=0.43cm]}
\newcommand{\ra}{\raisebox{-0.6ex}[0cm][0cm]}
\newcommand{\e}{\includegraphics}

\newcommand{\non}{\nonumber\\}

\begin{document}

\title{Dielectric, electromechanical and elastic properties of K$_{1-x}$(NH$_4$)$_x$H$_2$PO$_4$ compounds}

\author{
R.Levitskii, I.Zachek, L.Korotkov, D.Likhovaja, A.Vdovych,
\\S.Sorokov, Z.Trybula,  Sz.Los}

\date{}

\maketitle

 \begin{abstract}
 We present the results of the thermodynamic theory of piezoelectric ferroelectrics and of the phenomenological theory of thermodynamic characteristics
 of the KH$_{2}$PO$_{4}$ type ferroelectrics. Available experimental data for the dielectric, piezoelectric, and elastic characteristics of the KH$_{2}$PO$_{4}$
 type ferroelectrics and NH$_4$H$_{2}$PO$_{4}$ type antiferroelectrics are described using the proposed microscopic theory of these crystals.
 Using the results of the thermodynamic theory and experimental data we calculate the dielectric, piezoelectric, and elastic characteristics of the
  K$_{1-x}$(NH$_4$)$_x$H$_2$PO$_4$ type systems.
  \end{abstract}

\sloppy
\section{Introduction}

Since the 1990-ies, large attention has been paid to investigations of the
K$_{1-x}$(NH$_4$)$_x$H$_2$PO$_4$ type systems, undergoing the phase transitions
to the proton glass phase at low temperatures. Theoretical descriptio of the thermodynamic and dielectric
properties of these compounds is a complicated and unsolved problem of
statistical physics. Description of their dynamic properties within a microscopic approach
is particularly interesting. Experimental studies and theoretical description
of the temperature curves of real and imaginary parts of the dynamic dielectric permittivity tensor at different
frequencies, especially the low-temperature curves of the imaginary parts at low frequencies, are very important.

 In \cite{Vdovych111,Vdovych603,VdovychJPS}
 a cluster theory of static characteristics of model proton
glasses with an arbitrary range of competing interactions has been
proposed. In \cite{Vdovych523,cmp13706,Vdovych101,ferro43,izv1268,icmp1113e} a cluster theory of the thermodynamic and dynamic characteristics
of the  Rb$_{1-x}$(NH$_4$)$_x$H$_2$PO$_4$ type system has been proposed. It has been shown that at the proper choice of the model parameters this theory
 yields a satisfactory quantitative description of experimental data for these systems. Inconsistency of different experimental data was also revealed.

In \cite{Tu93,He91} it has been indicated that the piezoelectric coupling the
K$_{1-x}$(NH$_4$)$_x$H$_2$PO$_4$ type systems can exist. Unfortunately, in
\cite{Vdovych523,cmp13706,Vdovych101,ferro43,izv1268,icmp1113e} this coupling
was not taken into account.

In the paraelectric phase the ferroelectric compounds of the  MD$_2$XO$_4$ type (M = K, Rb, ND$_4$; X = P, As) crystallize in the  ${\bar 4}\cdot
 m$ class of tetragonal singony (the space group $I{\bar 4}2d$ with non-centrosymmetric point group
 $D_{2d}$). These crystals are piezoelectric in both phases  (paraelectric and ferroelectric or paraelectric and antiferroelectric), which essentially affects
 the behavior of their  physical characteristics.

Description of the dielectric properties of the MD$_{2}$XO$_{4}$
type ferroelectrics within the framework of the conventional
proton ordering model (see
\cite{16x,133x,17x,3lbm,471x,5lbm,uni2009}) was restricted to the
static limit and to the high-frequency relaxation. Attempts to
explore the piezoelectric resonance phenomenon within a model that
does not take into account the piezoelectric coupling are
pointless. The conventional proton ordering model does not
distinguish free and clamped crystals and is not able to reproduce
the effect of crystal clamping by high-frequency electric field.
This leads to an incorrect description of the temperature behavior
of the calculated polarization relaxation time and dynamic
dielectric permittivity of the  MD$_{2}$XO$_{4}$ type
ferroelectrics in the phase transition region.

Studies of the influence of the piezoelectric coupling on the
physical characteristics of the KH$_2$PO$_4$ type ferrroelectric
has been started in  \cite{0311U1}, where the Slater theory
\cite{0311U2} has been modified by taking into account the
splitting of the lowest ferroelectric level of the proton system
caused by the strain $\varepsilon_6$. More extensive results for
the deformed ferroelectrics of the KH$_2$PO$_4$ type were obtained
in
\cite{Izv705,0311U5,0311U6,JPS1701,Lis2003,Lis2004PSS,Lisnyj2004,0311U7,PCSS635,Lis2007,UFJ2008}.

In
\cite{Izv705,0311U5} a consistent microscopic formulation of the way with the strains of the different symmetries should be included into the proton
ordering model has been made. Later
\cite{0311U6,JPS1701,Lis2003,Lis2004PSS} all possible splittings of the proton configuration levels by the strains $\varepsilon_6$ were taken into account.
In \cite{0311U6} the phase transition in a deformed K(H$_{0,12}$D$_{0,88})_2$PO$_4$ crystal has been explored for the first time. The thermodynamic,
longitudinal dielectric, piezoelectric, and elastic characteristics of the crystal were calculated; their dependence on the stress $\sigma_6$ was studied.
 A thorough investigation of the thermodynamic and longitudinal dielectric, piezoelectric, and elastic characteristics of the  K(H$_{1-x}$D$_{x})_2$PO$_4$
  ferroelectrics was performed in  \cite{JPS1701}.

In \cite{Lisnyj2004}
a generalization of the proton ordering model  for the KH$_2$PO$_4$ type ferroelectrics was proposed, in order to explore the piezoelectric, dielectric, and elastic characteristics associated with the strains $\varepsilon_4$ and
 $\varepsilon_5$. The expressions for the transverse physical characteristics of these crystals
 in the paraelectric phase have been obtained and explored within the four-particle cluster approximation. By the proper choice
 of the model parameters, a good agreement between the theory and experiment for KH$_2$PO$_4$ and NH$_4$H$_2$PO$_4$.
A thorough investigation of transverse and longitudinal
characteristics of the NH$_4$H$_2$PO$_4$ and  ND$_4$D$_2$PO$_4$
antiferroelectric has been performed in  \cite{PCSS635}.

The relaxation phenomena in the KH$_2$PO$_4$  type ferroelectrics and NH$_4$H$_2$PO$_4$ type antiferroelectrics were explored within the modified proton
ordering model  with taking into account the
piezoelectric coupling  in \cite{cmp555,PCSS377,cmp275,cmp33705}.  The experimentally
 observed phenomena of crystal clamping by the high-frequency electric field, piezoelectric resonance,  and microwave dispersion were described.

Mixed ferroelectric-antiferroelectric (FE-AFE) crystals of the K$_{1-x}$(NH$_4$)$_x$H$_2$PO$_4$ type are a great example
of structural glasses. In these solid solutions a dipole glass state (DG) exists at low temperature and at certain
compositions. The T-x phase diagram of these compounds was explored in
\cite{483x,Gridnev91,Korotkov2004}. It is known that
the FE or AFE phase transitions are present in systems with ammonium content
 0.0<x<0.2 and 0.7<x<1.0, respectively. At
0.2<x<0.7 a glass-like behavior is observed at low temperatures.
At boundary compositions, the ferroelectric, paraelectric, and dipole glass
state coexist in a wide temperature range \cite{Gridnev91}.

These compounds, according to  \cite{Lines1977}, are of the KDP type and have a tetragonal structure ($I{\bar 4}2d$) at room temperature.
With lowering temperature the ferroelectric or antiferroelectric phase transitions are observed, attributed to the proton
ordering on the hydrogen bonds, connecting the neighboring PO$_4$ groups. In the paraelectric phase
the protons jump between two equilibrium positions on the bonds. Below the ferroelectric or antiferroelectric transition temperatures ($T_c$ or
$T_N$, respectively) a probability to find a proton in one of the two positions in the two-well potential on the bond increases.
Formation of the so-called "lateral" proton configurations leads to the antiferroelectric phase transition and to appearance
of the dipole moments in the  \textbf{ab} plane.
Appearance of the "up-down" configurations at the ferroelectric phase transitions leads to shifts of the heavy ions (P, O, and K) and gives rise to
 spontaneous polarization
(Ps) along the \textbf{c} axis. The ferroelectric and
antiferroelectric phases are orthorhombic, Fdd2 and
P2$_1$2$_1$2$_1$, respectively.

Even though the mixed crystals  K$_{1-x}$(NH$_4$)$_x$H$_2$PO$_4$
are well studied, their electromechanical properties are
practically not explored. In \cite{icmp1113e} an important role of
the piezoelectric coupling for the ferroelectric  crystals of the
KH$_2$PO$_4$ was shown. In \cite{NULP127,izv1414} a necessity for
a thorough investigation of the effects caused by the
piezoelectric coupling in these systems was underlined. Some
preliminary results of such investigations are presented in
\cite{icmp1113e}. In \cite{Korotkov1120,Korotkov52,Korotkov76} a
detailed experimental study of the dielectric, piezoelectric, and
elastic properties characteristics of these  materials is
reported.

In the earlier theoretical studies of the $Rb_{{\rm x}}
(NH_{4})_{{\rm 1-x}} H_{{\rm 2}} PO_{{\rm 4}} $ type mixtures, either only short-range \cite{Matsushita85} or only long-range \cite{Pirc87,Banerjee2003}
interactions were taken into account. However, in these crystals, the $PO_{4} $ (AsO$_{4} $) groups and their random surrounding by the
 $Rb$ or $NH_{4} $ ions play a crucial role in formation of the energy levels of the system
 and in creation of the random internal field. At the same time, an important role in formation of the ferroelectric and
 antiferroelectric structures is played by the long-range interactions.

Two types of random interactions in simple Ising-type systems with
pair interactions were taken into account within the cluster
approach in  \cite{Vdovych111,Vdovych603,VdovychJPS}. In
\cite{Vdovych523,Vdovych101,cmp13706,ferro43,izv1268} within the
four-particle approximation the temperature curves of
polarization, Edwards-Anderson parameters, longitudinal and
transverse permittivities of the $Rb_{{\rm x}} (NH_{4})_{{\rm
1-x}} H_{{\rm 2}} PO_{{\rm 4}}$ type system have been
satisfactorily described. However, for the transitional
composition ranges and in the phase transition temperature ranges
some qualitative and quantitative discrepancies between the theory
and experiment remain. We believe that to remove these
discrepancies one has to take into account the piezoelectric
coupling and use a better approximation for the fluctuations of
the random long-range interactions. In our previous papers we took
into account only the average values of the long-range
interactions. In this paper, for a simple model with random
short-range and long-range interactions and random internal field,
we take into account also the piezoelectric coupling. A useful
information about the mechanism of the phase transitions in the
  $Rb_{{\rm x}} (NH_{4})_{{\rm 1-x}} H_{{\rm 2}} PO_{{\rm
 4}}$ type systems can be obtained from experimental and theoretical studies of the
 influence of the external electric fields, hydrostatic and uniaxial pressures on the system. Correct calculations
 in this case are possible only with taking into account the piezoelelctric coupling.

In the present paper we consider in detail the results of the
thermodynamic theory of piezoelectrics and the results of the
phenomenological theory of the KH$_2$PO$_4$ type ferroelectrics.
The main theoretical results for the dielectric, piezoelectric,
elastic, and dynamic characteristics of the KH$_{2}$PO$_{4}$ type
ferroelectrics and NH$_4$H$_{2}$PO$_{4}$ type antiferroelectris
are also presented. Using the obtained experimental data, within
the thermodynamic theory the results for the dielectric,
piezoelectric, and elastic characteristics of the mixed
K$_{1-x}$(NH$_4$)$_x$H$_2$PO$_4$ systems are obtained.

\section{\large Piezoelectric properties of the crystals}
All ferroelectrics are also piezoelectric, at least in the ferroelectric phase \cite{143x}, because the ferroelectric crystals
are polar in the single domain state, or each domain is polar in the multi-domain state.
A multi-domain crystal as a whole can be either piezoelectric or non-piezoelectric. If this crystal is piezoelectric in the paraelectric phase,
that is, it is non-centrosymmetric, then, in accordance with the principle of recovering the paraelectric symmetry group after splitting into domains, the
crystal is piezoelectric in the ferroelectric phase in a multi-domain state.

The piezoelectricity is the phenomenon, when the mechanical stresses or strains induce
a linearly proportional to them electric polarization, or vice versa.
External influence on the crystal is described by intensive parameters, such as electric field $E_i$, mechanic stress $\sigma_j$,
temperature $T$. The internal state is described by the extensive parameters: polarization $P_i$, strains $\varepsilon_j$, entropy $S$.
The physical characteristics are determined by the relations between the external field
and the changes in its internal state. Hence, all characteristics is determined by the relations between two of the following parameters $E_i$, $P_i$,
$\sigma_j$, $\varepsilon_j$, $T$, and $S$.

Three of these characteristics are called the main ones; they relate two electric, two mechanic, and two thermal parameters. Those are
the dielectric susceptibility of mechanically free ($\sigma = const$) and clamped ($\varepsilon = const$) crystals  $\chi_{ii}^{\sigma, \varepsilon}$,
compliances $s_{ij}^{P, E}$, and molar specific heat $C_p$ at constant pressure
\setcounter{equation}{0}
\renewcommand{\theequation}{2.\arabic{equation}}
\bea
\chi_{ii}^{\sigma, \varepsilon} = \left( \frac{\partial P_i}{\partial E_i} \right)_{\sigma,\varepsilon}, \quad
s_{jj}^{P,E} = \left( \frac{\partial \varepsilon_j}{\partial \sigma_j} \right)_{P,E}, \quad
C_{p} = \left( \frac{\partial S}{\partial T} \right)_{p},
\eea
as well as the inverse dielectric susceptibility $k_{ii}^{\sigma,\varepsilon}$ and elastic constants $c_{ij}^{P,E}$:
\bea
k_{ii}^{\sigma, \varepsilon} = \left( \frac{\partial E_i}{\partial P_i} \right)_{\sigma,\varepsilon},
\quad
c_{jj}^{P,E} = \left( \frac{\partial \sigma_j}{\partial \varepsilon_j} \right)_{P,E}.
\eea
Here the indices denote the parameter, which should be kept constant during measurements.
Other characteristics are of a mixed nature and called conjugate.

In a ferroelectric piezoelectric crystal, the changes in the mechanical parameters $\sigma_j$ and $\varepsilon_j$
cause the changes in the electric parameters $E_i$ and $P_i$. This is the direct piezoelectric effect. At the converse
piezoelectric effect the changes in $E_i$ and $P_i$ induced the changes in $\sigma_j$ and $\varepsilon_j$.
The piezoelectric effect is characterized by the four piezoelectric moduli
\begin{enumerate}
\item[-] the coefficient of piezoelectric strain $d_{ij}$;
\item[-] the coefficient of piezoelectric voltage $e_{ij}$;
\item[-] the constant of piezoelectric voltage $h_{ij}$;
\item[-] the constant of piezoelectric strain $g_{ij}$.
 \end{enumerate}

These coefficient are determined as
\be
d_{ij} = \left( \frac{\partial P_i}{\partial \sigma_j} \right)_E, \,\,
e_{ij} = \left( \frac{\partial P_i}{\partial \varepsilon_j} \right)_E, \,\,
h_{ij} = -\left( \frac{\partial E_i}{\partial \varepsilon_j} \right)_P, \,\,
g_{ij} = -\left( \frac{\partial E_i}{\partial \sigma_j} \right)_P,
\ee
for the direct piezoeffect, and
\be
d_{ij} = \left( \frac{\partial \varepsilon_j}{\partial E_i} \right)_P, \,\,
e_{ij} = -\left( \frac{\partial \sigma_j}{\partial E_i} \right)_P, \,\,
h_{ij} = -\left( \frac{\partial \sigma_j}{\partial P_i} \right)_E, \,\,
g_{ij} = \left( \frac{\partial \varepsilon_j}{\partial P_i} \right)_E.
\ee
for the converse piezoeffect.
We dropped the index $T$ that indicated the fact that these characteristics are isothermal.

Hence, in the direct piezoeffect, the piezoelectric coefficient $d_{ij}$ determines the polarization of a clamped crystal
at given applied mechanical stress; the coefficient $e_{ij}$ determines the polarization caused by the strain;
the constant $h_{ij}$ determines the voltage in the open circuit at the given strain; the constant $g_{ij}$ determines
the voltage in the open circuit at the given stress. In the converse piezoeffect,
the coefficient $d_{ij}$ determines the strains in a free crystal at the given applied electric field;
the coefficient $e_{ij}$ determines the stress that has to be applied to a crystal to keep it undeformed at the given applied electric field;
the constant $h_{ij}$ determines the mechanical stress induced by polarization;
the constant $g_{ij}$ determines the strain induced by polarization.

To describe the piezoelectric effect, we use the free energy  $F$, which the function of
$P_i$, $\varepsilon_j$, $T$ ($i=1,2,3$, $j=4,5,6$) \cite{719x,138x,139x,142x,143x,17x,352x,147x,151x}. Hence,
\be
dF = E_i dP_i + \sigma_jd\varepsilon_j - SdT.
\ee
Then
\be
\sigma_j = \frac{\partial F}{\partial \varepsilon_j}, \quad
E_i = \frac{\partial F}{\partial P_i}, \quad  dS = - \frac{\partial F}{\partial T}.
\ee

Expanding the mechanical stress $\sigma_j$ and electric field $E_i$ at the isothermal conditions ($T = const$),
we obtain the following equations of the piezoelectric effect
\be
\varepsilon_j = c_{ij}^P \varepsilon_j - h_{ij} P_i,
\ee
\be
E_i = - h_{ij} \varepsilon_j + k_{ii}^\varepsilon P_i.
\ee

The second group of equations can be obtained from the Gibbs' function, which depends on $E_i$, $\sigma_j$, $T$:
\be
dG = - P_i dE_i - \varepsilon_j d\sigma_j - SdT.
\ee
Then
\be
\varepsilon_j = - \frac{\partial G}{\partial \sigma_j}, \quad
P_i = - \frac{\partial G}{\partial E_i}, \quad
dS = - \frac{\partial G}{\partial T}.
\ee
Expanding the  strains $\varepsilon_j$ and polarization $P_i$, we obtain the second pair of the piezoeffect equations
\be
\varepsilon_j = S_{ij}^E \sigma_j + d_{ij} E_i,
\ee
\be
P_i = d_{ij} \sigma_j + \chi_{ii}^\sigma E_i.
\ee

Using the elastic Gibbs' function
\be
dG_1 = - \varepsilon_j d\sigma_j + E_i dP_i - SdT
\ee
and electric Gibbs' function
\be
dG_2 = \sigma_j d\varepsilon_j - P_i dE_i - SdT
\ee
we get the third and the fourth pairs of the isothermal piezoeffect equations
\be
\varepsilon_j = S_{ij}^P \sigma_j - g_{ij}P_i,
\ee
\be
E_i = - g_{ij}\sigma_j + k_{ii}^\sigma P_i;
\ee
\be
\sigma_j = C_{jj}^E \varepsilon_j - e_{ij}E_i,
\ee
\be
P_i = e_{ij} \varepsilon_j + \chi_{ii}^\varepsilon E_i.
\ee

We can also find the relations between the dielectric, piezoelectric, and elastic characteristics
\bea
&& d_{ij} = g_{ij} \chi_{ii}^\sigma = e_{ij} s_{jj}^E, \nonumber\\
&& e_{ij} = h_{ij} \chi_{ii}^\varepsilon = d_{ij} c_{ij}^E,\nonumber\\
&& h_{ij} = e_{ij} k_{ii}^\varepsilon = g_{ij} c_{jj}^P, \\
&& g_{ij} = d_{ij} k_{ii}^\sigma = h_{ij}s_{jj}^P, \nonumber\\
&& \chi_{ii}^\sigma = \chi_{ii}^\varepsilon + e_{ij}d_{ij}; \quad
k_{ii}^\sigma = k_{ii}^\varepsilon - h_{ij} g_{ij},\\
&& c_{jj}^P = c_{jj}^E + e_{ij} h_{ij}, \quad
s_{jj}^E = s_{jj}^P + h_{ij} g_{ij}.
\eea

The four piezoelectric coefficients are not independent: knowing
one of them and using elastic or dielectric characteristics, we
can calculate the rest of the piezoelectric coefficients. The
piezoelectric coefficients correspond to different measurement
conditions; these conditions are also determined by the crystal
state. Thus,  the dielectric characteristics are different for the
mechanically free and clamped states of the crystal. On the other
hand, the elastic characteristics are different for open or short
circuited crystals. Mechanical and electrical conditions at
measurements should be also taken into account at transitions
between the piezoelectric coefficients.

A mechanically clamped crystal cannot be deformed $(\varepsilon=0)$. This state is indicated
by the index $\varepsilon$ at $\chi_{ii}$ and $\kappa_{ii}$. In a mechanically free crystal
the stresses are absent during the measurements. This state in indicated by the index $\sigma$ at $\chi_{ii}$ and $\kappa_{ii}$.
Sometimes, the crystal states of constant strains $(\varepsilon_{j}=const)$ or constant stresses $(\sigma_{j}=const)$
are considered, but such a consideration gives no principally new result.

In ferroelectrics the quantities measured at constant stresses or
at constant strains can be essentially different because of the
elastic contribution into the compliances. Measurements for a
clamped crystal are usually performed using the dynamical method
at frequencies far above the main resonant frequency of the
sample, whereas the measurements for a free crystal are performed
at low frequencies or quasi-statically.

In a short-circuited crystal (electrically free) all the surface
is at the same potential  $(E=0)$. This state is often realized
when the pair of electrodes, which is short-circuited, is the one
covering the faces at which the piezoelectric polarization arises
during measurements of $s$ and $c$.

Electrically open crystal is completely isolated. This does not necessarily mean $P=0$.
In the case of a isolated plate the condition $P=0$ is fulfilled only in the case, when
the polarization is perpendicular to the crystal faces. If $P=0$, the crystal is electrically clamped.
The state with $E=0$ and $P=0$ is marked by the corresponding indices at $s$ and $c$.

In absence of the electric field $(E_{i}$= 0) applied to a free crystal  $(\sigma_{j}$= 0), and if the spontaneous strains are neglected
$(\varepsilon_{j}$= 0) we have $F=G=G_{1}=G_{2}$. For a free crystal in external field, if the piezoeffect is neglected
 $ F=G=G_{1}+\vec{P}\vec{E}=G_{2}+\vec{P}\vec{E}$.

To describe the ``clamped'' ferroelectrics it is convenient to use the variables $P_{s}$ and $\varepsilon_{j}$, that is, to use
the free energy $F(P_{i},\varepsilon_{j},T)$. To describe the free crystals it is convenient to
use the elastic Gibbs' function $G_{1}(P_{i},\sigma_{j},T)$.

The susceptibility $\chi_{33}^\sigma$ increases with decreasing temperature by the hyperbolic law and reaches about $10^5$ at the transition point.
Therefore, according to the expression $d_{36} = g_{36} \chi_{33}^\sigma$, the piezomodule $d_{36}$, which relates
the polarization along the $c$ axis to the mechanical stress $\varepsilon_6$, also has an anomaly at the Curie temperature.

The compliance $s_{66}^E$ also has an anomalous temperature behavior, since $s_{66}^E = s_{66}^P + g_{36} d_{36}$.
The origin of this anomaly is the following. In experimental conditions $E=const$ means $E=0$. The sample faces perpendicular
to $c$ are electroded and short-circuited. If ones apply the shear stress $\sigma_6$ to the short-circuited crystal,
the polarization along the  $c$ axis arises. Due to the converse piezoeffect, this also increases the
shear strain $\varepsilon_6$. This effect is particularly strong near the Curie point, where the piezomodule $d_{36}$ increases anomalously.
Therefore, the compliance of the crystal with respect to
 the shear stress $\sigma_6$, namely $s_{66}^E$, should increase anomalously near the Curie temperature.  At $E \neq const$
 (an ``isolated'' crystal) the polarization $P_3$, induced by the strain $\sigma_6$, will create
  a depolarization field $E_3 = - 4\pi P_3$, which decreases the polarization practically to zero. In this case $P=0$. Since the polarization is absent,
  then there is no strain  $\varepsilon_6$ either, and the compliance  $s_{66}^P$ has no anomaly at the Curie point.

\section{Phenomenological theory of the thermodynamic characteristics}
The KH$_2$PO$_4$ crystal undergoes the first order phase transition. Its spontaneous polarization $P_{s3}$ is proportional to the
order parameter $\eta^{(1)}$; in the ferroelectric phase the polarization is accompanied by the spontaneous strain $\varepsilon_{s6}$.
Expansion of the crystal free energy in series over the order parameter in absence of external electric field or mechanical stress contains only
terms of the even order and reads
\setcounter{equation}{0}
\renewcommand{\theequation}{3.\arabic{equation}}
\be
F_1(P_{s3}, \varepsilon_{s6}, T ) = F_0 + \frac12 a^\varepsilon(T) P_{s3}^2 + \frac14 b^\varepsilon P_{s3}^4 + \frac16 c^\varepsilon P_{s3}^6.
\ee
It should be supplemented with the elastic energy
%%% 1.7
\be
F_{el} (P_{s3}, \varepsilon_{s6}, T) = \frac12 c_{66}^P \varepsilon_{s6}^2 - h_{36} P_{s3} \varepsilon_{s6}.
\ee

The quantities $P_{s3}$ and  $\varepsilon_{s6}$ can be found from equations, following from the equilibrium conditions
%%% 1.8
\be
\frac{\partial F}{\partial P_{s3}} = 0, \quad \frac{\partial F}{\partial \varepsilon_{s6}} = 0,
\ee
where $F = F_1 + F_{el}$.

The coefficient $a^\varepsilon(T)$ should turn to zero at the stability limit of the paraelectric phase $T_0$, where $T_0$ is the
Curie-Weiss temperature. Therefore, in the vicinity of the stability limit $a(T)$ can be expanded in $(T-T_0)$, retaining only the linear term
\[
a^\varepsilon(T) = a^{\varepsilon'}(T-T_0), \quad a^{\varepsilon'} =
\left( \frac{\partial a}{\partial T} \right)_{T=T_0} > 0.
\]
In the case of the first order phase transition $b^\varepsilon < 0$; we replace $b= - {\bar b}$. The coefficients
$a, {\bar b}, c$ ц $F(P_{53}, \varepsilon_{56}, T)$ have the index ``$\varepsilon$'',  since they are
taken at constant strain and are positive.

From (1.24) we obtain the expressions for the spontaneous polarization of KH$_2$PO$_4$, which corresponds to the
free energy minimum
%%% 1.9
\be
P_{s3} = \left\{ \frac{{\bar b}^\varepsilon}{2c^\varepsilon} \left[ 1 + \sqrt{1 - \frac{4c^\varepsilon}{{\bar b}^{\varepsilon2}} \left( a^{\varepsilon'} (T-T_0) -
\frac{h_{36}^2}{c_{66}^P}\right)} \right] \right\}^{\frac12},
\ee
and for the spontaneous strain
%%% 1.10
\be
\varepsilon_{s6} = \frac{h_{36}}{c_{66}^P} \left\{ \frac{{\bar b}^\varepsilon}{2c^\varepsilon}
\left[ 1 + \sqrt{1 - \frac{4c^\varepsilon}{{\bar b}^{\varepsilon2}} \left( a^{\varepsilon'} (T-T_0) -
\frac{h_{36}^2}{c_{66}^P}\right)} \right] \right\}^{\frac12}.
\ee

In presence of external fields the polarization is $P_3 = P_{s3} + P_{i3}$, and the strain is $\varepsilon_6 = \varepsilon_{s6} + \varepsilon_{i6}$.
If the external electric field $E_3$ and mechanical stress  $\sigma_6$ are applied, we use the elastic Gibbs' function
%% 1.11
\be
G_1(P_3,\sigma_6,T) = G_{10} + \frac12 a^\sigma P_3^2 + \frac14 b^\sigma P_3^4 + \frac16 c^\sigma P_3^6 + \frac12 s_{66}^P\sigma_6^2 + g_{36}P_3 \sigma_6.
\ee

The first order phase transition in zero field takes place, when both $G_1$ and its first derivative with respect to $P_3$
are simultaneously equal to zero at non-zero $P_3$. From the condition $\left( \frac{\partial G_1}{\partial \sigma_6} \right) = 0$
we find that $\sigma_6 = - \frac{g_{36}}{s_{66}^P}P_3$; then the two conditions%% 1.12, 1.13
\bea
&& \left[ \frac{a^{'\sigma}}{2} (T-T_0) - \frac12 \frac{g_{36}^2}{s_{66}^P} \right] -
\frac{{\bar b}^0}{4}P_3^2 - \frac{c^\sigma}{6}P_3^4 = 0, \\
&& a^{'\sigma}(T-T_0) - \frac{g_{36}^2}{s_{66}^P} - {\bar b}^\sigma P_3^2 + c^\sigma P_3^4 = 0
\eea
are satisfied simultaneously \cite{147x}. The solution of (1.28) reads
%% 1.14
\be
T = T_c = T_0 + \frac{3}{16} \frac{{\bar b}^{\sigma 2}}{a^{'\sigma} c^\sigma} + \frac{1}{a^{'\sigma}}
\frac{g_{36}^2}{s_{66}^P}.
\ee
The expression for  $T_c$ without the last term is presented in  \cite{147x,17x}.

Substituting (1.30) into (1.29), we find spontaneous polarization at $T_c$:
%% 1.15
\be
P_{3c}^2 = \frac34 \frac{\bar b}{c} + \frac{4}{\bar b} \frac{g_{36}}{s_{66}^P}.
\ee

From the conditions of equilibrium
\[
\left( \frac{\partial G_1}{\partial P_3}\right) _{\sigma=0} =0, \quad
\left( \frac{\partial G_1}{\partial \sigma_6}\right) = \varepsilon_6
\]
we obtain expressions for polarization
%% 1.16
\be
P_3 = \left\{ \frac{{\bar b}^\sigma}{2c^\sigma} \left[ 1 + \sqrt{1 - \frac{4c^\sigma}{{\bar b}^{\sigma 2}} \left( a^{\sigma'} (T-T_0) -\frac{g_{36}^2}{s_{66}^P}\right)} \right] \right\}^{\frac{1}{2}}
\ee
and strain
%% 1.17
\be
 \varepsilon_6 = s_{66}^P \sigma_6 + g_{36} \left\{ \frac{{\bar b}^\sigma}{2c^\sigma} \left[ 1 + \sqrt{1 - \frac{4c^\sigma}{{\bar b}^{\sigma 2}} \left( a^{\sigma'} (T-T_0) -\frac{g_{36}^2}{s_{66}^P}\right)} \right] \right\}^{\frac{1}{2}}.
\ee

The expression for the $P_3$ is real at all temperatures $T < T_0^-$, where $T_0^-$ is the stability limit of the ferroelectric phase
\[
T_0^- = T_0 + \frac{{\bar b}^{\sigma 2}}{4a^{'\sigma}} \left( 1 + \frac{g_{36}^2}{s_{66}^P} \right).
\]
At $T = T_0^-$ the polarization $P_3$ is finite.

To find the static dielectric permittivity  $\varepsilon_{33}^\sigma$ we use the conditions
%% 1.18
\be
\left( \frac{\partial G_1}{\partial P_3}\right) _{\sigma=0} =E_3, \quad
\frac{\partial G_1}{\partial \sigma_6} = \varepsilon_6.
\ee

As a result
%% 1.19
\be
\left( a^\sigma - \frac{g_{36}^2}{s_{66}^P} \right) P_3 - {\bar b}^\sigma P_3^3 + c^\sigma P_3^5 = E_3.
\ee

Since $E_3$ is small, Eq.~(1.35) can be linearized. For that the polarization can be presented as
%% 1.20
\be
P_3 = P_{s3} + P_{i3},
\ee
where $P_{i3}$ is small. Using (1.35), neglecting the terms with $\sim P_{i3}^2$ and higher, and taking into account the
definition of $P_{i3} = \frac{\varepsilon_{33}^\sigma -1}{4\pi}E_3$, we find and expression for the static dielectric
permittivity
\[
\varepsilon_{33}^\sigma = 1 + 4\pi \chi_{33}^\sigma,
\]
where
%% 1.21
\bea
&& \chi_{33}^\sigma = \frac{1}{a^{'\sigma}(T-T_0) - \frac{g_{36}^2}{s_{66}^P} - 3{\bar b}^\sigma P_{s3}^8 + 5c^\sigma P_{s3}^4} = \nonumber\\
&& = \left \{ \frac{{\bar b}^{\sigma 2}}{c^\sigma} \sqrt{1 - \frac{4c^\sigma}{{\bar b}^{\sigma 2}} \left[ a^{'\sigma} (T-T_0) -\frac{g_{36}^2}{s_{66}^P}\right]} \times    \right. \\
&& \left. \times \left\{ 1 + \sqrt{1 - \frac{4c^\sigma}{{\bar b}^{\sigma 2}} \left[ a^{'\sigma} (T-T_0) -\frac{g_{36}^2}{s_{66}^P}\right]} \right\} \right\}^{-1}. \nonumber
\eea

The stability limit of the ferroelectric phase is determined by zero of this expression and coincides
with  $T_0^-$. Above the transition temperature
\[
\chi_{33}^\sigma = \frac{C_{\chi}}{T- T_0^+},
\]
where $C_{\chi} = \frac{1}{a^{'\sigma}}$  is the Curie-Weiss constant; $T_0^+ = T_0 + \frac{1}{a^{'\sigma}} \frac{g_{36}^2}{s_{66}^P}$. For
 KH$_2$PO$_4$ $C_\varepsilon = 4\pi C_{\chi} = 3250$~K. The Curie-Weiss law is
well obeyed within the temperature range of  50~K above $T_0^+$.

The susceptibility $\Bigl( \chi_{33}^\sigma \Bigr)^{-1}$ is finite; at $T = T_c$ is has a discontinuity;
and the stability limits of the paraelectric $(T_0^+)$ and ferroelectric $(T_0^-)$ phases do not coincide, which is typical for the
first order phase transitions.

From (1.15) at $\sigma_6 = 0$ we get $P_3 = \frac{\varepsilon_6}{g_{36}}$. Substituting this into (1.35), we have
%% 1.22
\be
\left( a^\sigma - \frac{g_{36}^2}{s_{66}^P} \right) \frac{\varepsilon_6}{g_{36}} - \frac{b^\sigma}{g_{36}^3} \varepsilon_6^3 + \frac{c^\sigma}{g_{36}^5} \varepsilon_6^5 = E_3.
\ee

We present the strain $\varepsilon_6$ in the form
%% 1.23
\be
\varepsilon_6 = \varepsilon_{s6} + \varepsilon_{i6}.
\ee
Let us substitute (1.39) into (1.38) and neglect the terms with $\sim \varepsilon_{i6}^2$ and higher.
In the result we obtain an expression for the coefficient of piezoelectric strain
%% 1.24
\bea
&& d_{36} = \frac{\varepsilon_{i6}}{E_3} =
\frac{g_{36}}{\left( a^\sigma - \frac{g_{36}^2}{s_{66}^P} \right) - \frac{3{\bar b}^\sigma}{g_{36}^2}\varepsilon_{56}^2 + \frac{5{\bar c}^\sigma}{g_{36}^4}\varepsilon_{56}^4} = \nonumber\\
&& = \frac{g_{36}}{a^{'\sigma}(T-T_0) - \frac{g_{36}^0}{s_{66}^P} - 3{\bar b}^\sigma P_{s6}^2 + 5c^\sigma  P_{s6}^4} = g_{36}\chi_{33}^\sigma.
\eea

In the paraelectric phase
%% 1.25
\be
d_{36} = \frac{B}{T - T_0^+},
\ee
where $B = \frac{g_{36}}{a^{'\sigma}}$ is the Curie-Weiss constant.

The major advantage of the thermodynamic theory is its
mathematical simplicity, wide range of applications, and a possibility to find the relations between various macroscopic
parameters of the ferroelectrics.
Its limitations are caused by its purely macroscopic nature, which excludes any microscopic insight into the transition origin or atomic processes
associated with the ferroelectricity. In fact, this theory is phenomenological.

\section{Model Hamiltonian of the KH$_{2}$PO$_{4}$ and NH$_{4}$H$_{2}$PO$_{4}$ crystals}

\begin{figure}[h!]
\begin{center}
  \includegraphics[scale=0.5]{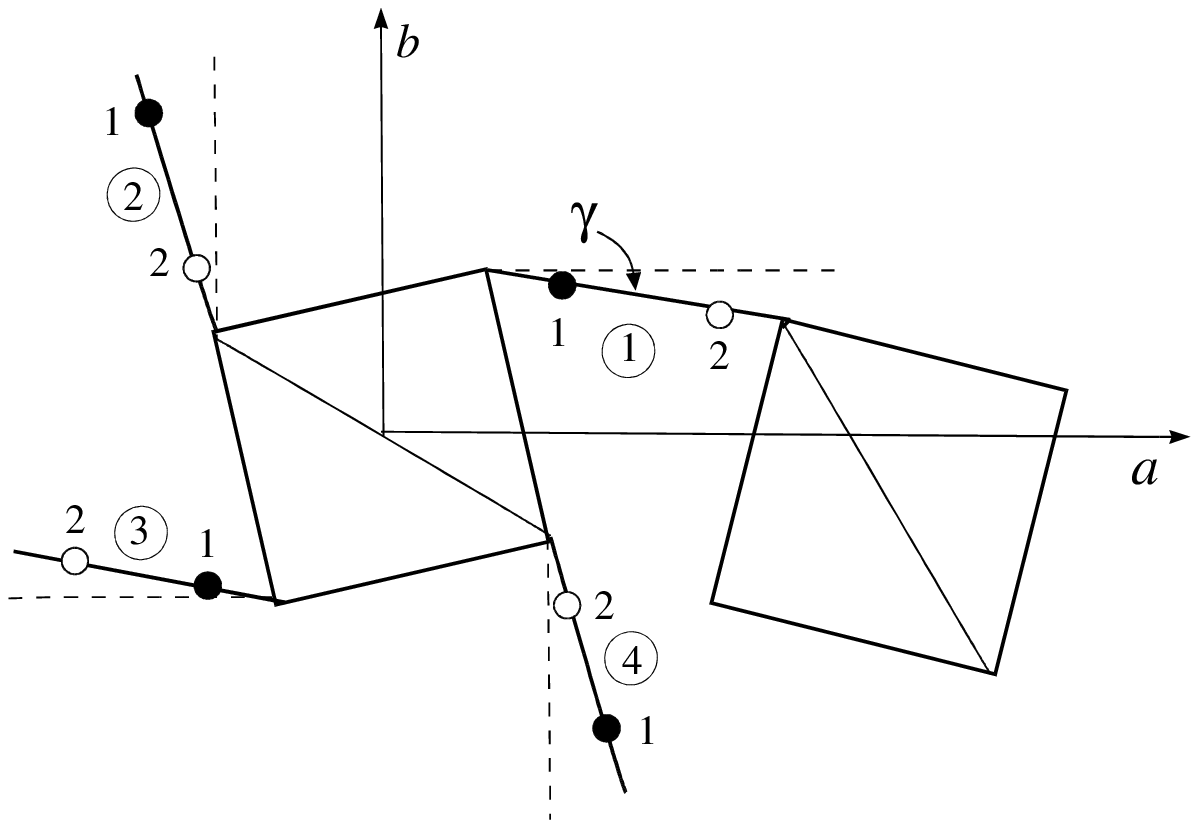}~~~~  \includegraphics[scale=0.65]{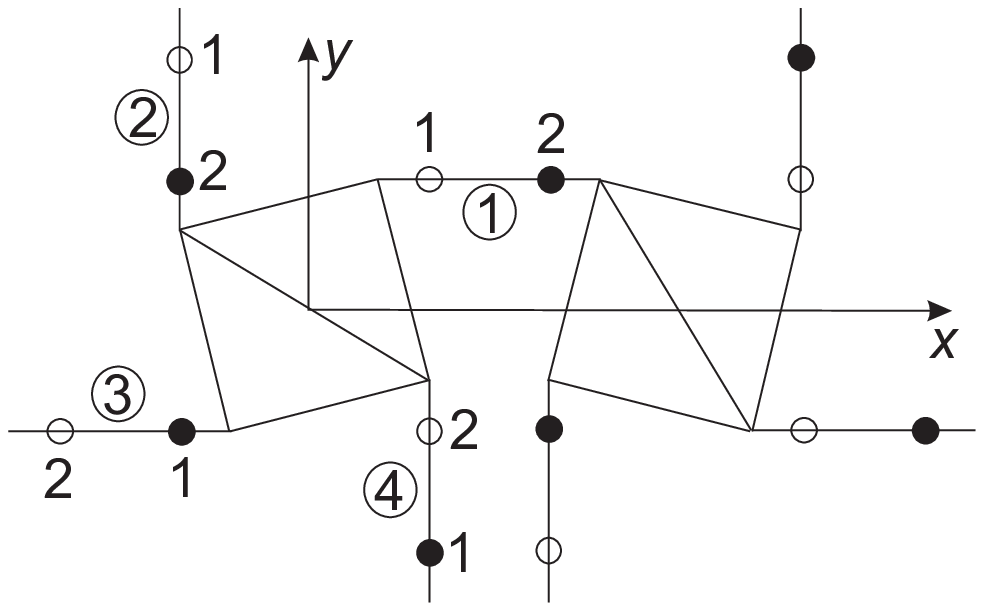}\\
   a ~~~~~~~~~~~~~~~~~~~~~~~~~~~~~~~~~~~~~~~~~~~~~~~~~~ b
\end{center}
\caption[]{A primitive cell of the KDP (a) and ADP (b) crystals. The numbers \raisebox{-0.70ex}[0cm][0cm]{\eeee{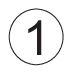},
\eeee{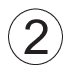}, \eeee{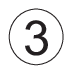}, \eeee{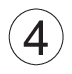}} indicate the hydrogen bonds: 1, 2 are possible equilibrium positions of the
protons. Two of possible proton configurations (ferroelectric (a) and antiferroelectric (b)) are shown.} \label{fig1}

\end{figure}

\setcounter{equation}{0}
\renewcommand{\theequation}{4.\arabic{equation}}
We shall consider a system of protons moving on the  O-H...O bonds
in  KH$_{2}$PO$_{4}$ (KDP) and NH$_{4}$H$_{2}$PO$_{4}$ (ADP) crystals.
 The
primitive cell of the Bravais lattice of these crystals consists of
two neighboring tetrahedra PO$_4$ along with four hydrogen bonds
attached to one of them (the "A'' type tetrahedron). The hydrogen
bonds attached to the other tetrahedron ("B'' type) belong to four
surrounding it structural elements
(fig.~\ref{fig1}).

The calculations of the physical characteristics of the KDP and ADP crystals are performed in the four-particle cluster approximation
for the  short-range interactions and the mean field approximation for the
long-range interactions in presence of external electric field $E_3$ along the crystallographic $c$ axis and mechanical stress $\sigma_6 = \sigma_{xy}$. In
absence of tunneling the system Hamiltonian reads
\setcounter{equation}{0}
\renewcommand{\theequation}{3.\arabic{equation}}
\bea && \hspace{-4ex} \hat H = NH^{0} + \frac12
\sum_{qq'}J_{ff'}(qq')
\frac{\langle\sigma_{qf}\rangle}{2}\frac{\langle\sigma_{q'f'}\rangle}{2}
+ \sum_{q}\hat H_{q,s,a}^{(4)}, \label{2.1} \eea
where $N$ is the total number of primitive cells; $\sigma_{qf}$ is the operator of the
z-component of the pseudospin, which eigenvalues $\sigma_{qf}=\pm 1$ correspond to the two equilibrium proton position in the
q-th cell on the  f-th bond.  The ``seed''
energy corresponds to the sublattice of heavy ions and does not depend explicitly
on the deuteron subsystem configuration. It is expressed in terms of the strain $\varepsilon_6$ and electric field
 $E_3$ and includes the elastic, piezoelectric, and dielectric contributions
\bea && \hspace{-4ex} H^{0} = \frac{
v}{2}c_{66}^{E0}\varepsilon_6^2 - { v}e_{36}^{0}
\varepsilon_6E_3 - \frac{
v}{2}\chi_{33}^{\varepsilon0} E_3^2. \label{2.2} \eea
where $v$ is the
primitive cell volume; $c_{66}^{E0} $, $e_{36}^0$, ${\chi }_{33}^{\varepsilon 0} $are the ``seed'' elastic constant,
piezoelectric coefficient, and dielectric susceptibility. They determine the temperature behavior of
the corresponding observable quantities at temperatures far from the phase transition.

The four-particle proton Hamiltonians $H_{q,s,a}^{(4)}$ read%
\bea && \hspace{-8ex} \hat H_{q,s}^{(4)} = \hat H_{q}^{(4)s}   -  \sum_{f=1}^4\frac{z_6}{\beta}
\frac{\sigma_{qf}}{2}, \label{2.3} \eea
\bea && \hspace{-8ex} \hat H_{q,a}^{(4)} = \hat H_{q}^{(4)a} -\! \frac{1}{\beta}x_{q} \!\left(-\frac{\sigma_{q1}}{2}
\!+\! \frac{\sigma_{q2}}{2} \!+\! \frac{\sigma_{q3}}{2} \!-\!
\frac{\sigma_{q4}}{2}\right)   -  \sum_{f=1}^4\frac{z}{\beta}
\frac{\sigma_{qf}}{2}, \label{2.3a} \eea

\bea && \hspace{-8ex} \hat H_{q}^{(4)s,a} = (
-\delta_{s6}\varepsilon_6 - 2\delta_{16}\varepsilon_6 )
\left(\frac{\sigma_{q1}}{2}\frac{\sigma_{q2}}{2}\frac{\sigma_{q3}}{2}
+ \frac{\sigma_{q1}}{2}\frac{\sigma_{q2}}{2}\frac{\sigma_{q4}}{2}
+
\frac{\sigma_{q1}}{2}\frac{\sigma_{q3}}{2}\frac{\sigma_{q4}}{2} +
\frac{\sigma_{q2}}{2}\frac{\sigma_{q3}}{2}\frac{\sigma_{q4}}{2}\right) + \nonumber\\
&& \hspace{-2ex} +2( \delta_{a4}\varepsilon_4 -
\delta_{14}\varepsilon_4 )
\left(\frac{\sigma_{q1}}{2}\frac{\sigma_{q2}}{2}\frac{\sigma_{q4}}{2}
- \frac{\sigma_{q2}}{2}\frac{\sigma_{q3}}{2}\frac{\sigma_{q4}}{2}
\right) + \label{2.3x}\\
&& \hspace{-2ex} + (V_{s,a} + \delta_{a6} \varepsilon_6)
\left(\frac{\sigma_{q1}}{2}\frac{\sigma_{q2}}{2} +
\frac{\sigma_{q3}}{2}\frac{\sigma_{q4}}{2}\right) +(V_{s,a} - \delta_{a6} \varepsilon_6)
\left(\frac{\sigma_{q2}}{2}\frac{\sigma_{q3}}{2} +
\frac{\sigma_{q4}}{2}\frac{\sigma_{q1}}{2}\right) +
\nonumber\\
&& \hspace{-2ex} + U_{s,a} \left( \frac{\sigma_{q1}}{2}
\frac{\sigma_{q3}}{2} + \frac{\sigma_{q2}}{2}
\frac{\sigma_{q4}}{2} \right) + \Phi_{s,a} \frac{\sigma_{q1}}{2}
\frac{\sigma_{q2}}{2} \frac{\sigma_{q3}}{2} \frac{\sigma_{q4}}{2}, \nonumber \eea
\bea &&  \hspace{-4ex} x_q = \beta (-\Delta_a e^{i{\bf k}^z{\bf a}_q} + 2\nu_a({\bf
k}^z) \eta^{(1)} e^{i{\bf k}^z{\bf a}_q}), \nonumber\\
&& \hspace{-4ex}  z_6 = \beta (-\Delta_{c} + 2\nu_c(0) \eta^{(1)} -
2\psi_6\varepsilon_6+\mu_3 E_3), \nonumber\\
&& \hspace{-4ex}  z_6 = \beta (-\Delta_{c} + 2\nu_c(0) \eta^{(1)z} -
2\psi_6\varepsilon_6+\mu_3 E_3),  \nonumber \eea
where
\bea
 && 4\nu_c(0) = J_{11}(0) + 2J_{12}(0) + J_{13}(0),   \nonumber\\
&&  4\nu_a^0({\bf k}^z) = J_{11}({\bf k}^z) - J_{13}({\bf k}^z), ~~~~ J_{ff'}({\bf k}^z) =
\sum\limits_{{\bf a}_q - {\bf a}_{q'}}J_{ff'}(qq')
e^{-i{\bf k}^z({\bf a}_q- {\bf a}_{q'})}, \nonumber \eea
and ${\bf k}^z = 1/2 ({\bf b}_1 + {\bf b}_2 + {\bf b}_3)$, ${\bf
b}_1$, ${\bf b}_2$, ${\bf b}_3$ are the vectors of the reciprocal lattice;
$e^{i{\bf k}^z{\bf a}_q} = \pm 1$; $\psi_{6}$ is the deformation potential;
$\Delta_{a}$ and $\Delta_{c}$ are the effective fields
exerted by the neighboring hydrogen bonds from
outside the cluster. $\mu_{3}$ is the effective dipole moment.

$\Delta_{s}^c$, ,
is determined from the self-consistency condition: the
mean values $\langle \sigma_{qf} \rangle$ calculated within the
four-particle and one-particle cluster approximations should
coincide.

In (\ref{2.3x})
\[ V_s = -\frac{w_1}{2}, \quad U_s = \frac{w_1}{2}-\varepsilon, \quad \Phi_s =
4\varepsilon - 8w + 2w_1,\]
\[
V_a = \frac12 \varepsilon' - \frac12 w_1', \qquad U_a = \frac12
\varepsilon' + \frac12 w_1', \qquad \Phi_a = 2\varepsilon' - 8w' +
2w_1',\]
 and $
\varepsilon=\varepsilon_a-\varepsilon_s$,
$w=\varepsilon_1-\varepsilon_s$, $w_1=\varepsilon_0-\varepsilon_s$, $\varepsilon'=\varepsilon_s-\varepsilon_a$,
$w'=\varepsilon_1-\varepsilon_a$, $w'_1=\varepsilon_0-\varepsilon_a$,
where $\varepsilon_s$, $\varepsilon_a$, $\varepsilon_1$,
$\varepsilon_0$ are the configurational energies of protons near the
PO$_4$ tetrahedra.

In absence of the external electric field or stress,
 we have the following equation for $\eta^{(1)}$
 \bea &&
 \eta_s^{(1)} = \langle \sigma_{q1} \rangle = \langle \sigma_{q2} \rangle
 = \langle \sigma_{q3} \rangle = \langle \sigma_{q4} \rangle
  =  \label{etas}\\
&&  ~~~~= \frac{1}{D_s}(\sinh (2z_6 + \beta \delta_{s6}\varepsilon_6) + 2b \sinh(z_6 - \beta
 \delta_{16}\varepsilon_6)), \nonumber\\
&&  D_s= \cosh (2z_6 + \beta \delta_{s6}\varepsilon_6) + 4b \cosh
 (z_6 - \beta \delta_{16}\varepsilon_6) + 2a \cosh \beta
 \delta_{a6}\varepsilon_6 + d, \nonumber
 \eea
 \bea
 && \hspace{-4ex} z_6 = \frac12 \ln \frac{1 + \eta_s^{(1)}}{1 - \eta_s^{(1)}}
 + \beta \nu_c\eta_s^{(1)} - \beta\psi_6\varepsilon_6, ~~ a = e^{-\beta \varepsilon}, ~~ b = e^{-\beta w}, ~~ d = e^{-\beta
 w_1}; \nonumber
 \eea
and
\bea && \eta_a^{(1)} = - \langle \sigma_{q1} \rangle = \langle \sigma_{q2}
\rangle = \langle \sigma_{q3} \rangle = - \langle \sigma_{q4}
\rangle = \frac{1}{D_a}(\sinh 2x + 2b' \sinh x),  \label{etaa}\\
 &&
D_a = a' + \cosh 2x + d' + 4b' \cosh x + 1, ~~~~ x = \frac12 \ln
\frac{1 + \eta_a^{(1)}}{1 - \eta_a^{(1)}} + \beta \nu_a ({\bf
k}^z) \eta_a^{(1)}.
\nonumber\\
&&
a' = e^{-\beta \varepsilon'}, ~~~~ b' = e^{-\beta
w'},~~~~
d' = e^{- \beta w'_1}.
\nonumber \eea

\section{Dielectric, piezoelectric, and elastic characteristics}

The dynamic characteristics of KDP and ADP
crystals in presence of the piezoelectric coupling to the strain  $\varepsilon_6$ will be explored using the proposed dynamic
model  based on a stochastic Glauber
model, with taking into account  dynamics of the strains using Newtonian equations of motion  \cite{cmp555,cmp275}.
 Using the method developed in these papers, we obtain the following expressions for the longitudinal dynamic susceptibilities
 of mechanically free KDP and ADP crystals
\setcounter{equation}{0}
\renewcommand{\theequation}{5.\arabic{equation}}
 \begin{eqnarray}
 && \chi_{33}^{\sigma}(\omega) = \chi_{33}^{\varepsilon}(\omega)+  {R_6(\omega)}
 \frac{e_{36}^2(\omega )}{c_{66}^E(\omega)}, \label{3.7}
 \end{eqnarray}
where \cite{cond2012}
 \be
 {R_6(\omega)} = 1+\sum\frac{64}{(2k+1)^{2}(2l+1)^{2}\pi^{4}}\frac{\omega^{2}}{(\omega_{kl}^{0})^{2}-\omega^{2}},
 %~~~~ k_{6} = \frac{\omega\sqrt{\rho}}{ \sqrt{c_{66}^E(\omega)} },
 \omega_{kl}^{0} = \sqrt{\frac{c_{66}^{E}(\omega_{kl}^{0})\pi^{2}}{\rho}[\frac{(2k+1)^{2}}{L_{x}^{2}}+\frac{(2l+1)^{2}}{L_{y}^{2}}]}.
 \label{k6}
 \ee

In (\ref{3.7}) the longitudinal dynamic susceptibility of mechanically clamped KDP crystals, the piezoelectric coefficient, and
the elastic constant
are
 \bea && \hspace{-2ex}
 \chi_{33}^{\varepsilon}(\omega) = \chi_{33}^{\varepsilon 0} +
 \frac{\beta\mu_3^2}{2v}F^{(1)}(\omega), \label{x33e} \\
&& \hspace{-2ex} e_{36} (\omega) = e_{36}^0 + \label{e36} \\
 && \hspace{-2ex}+
 \frac{\beta\mu_3}{v} \Bigl[ -\psi_6 F^{(1)}(\omega) +
 \delta_{s6}F_s^{(1)}(\omega) +  \delta_{16}F_1^{(1)}(\omega) -
 \delta_{a6}F_a^{(1)}(\omega)\Bigr],\non
 && \hspace{-2ex}  c_{66}^E( \omega) = c_{66}^{E0} + \frac{4\beta \psi_6}{v D_s}f_s + \frac{2\beta}{v D_s^2} (- \delta_{s6}M_{s6} + \delta_{16}M_{16} +
 \delta_{a6}M_{a6})^2 +  \label{c66E} \\
 && \hspace{-2ex} + \frac{4\beta
 \psi_6}{v} \Bigl[ - \psi_6 F^{(1)}( \omega) +
 \delta_{s6}F_s^{(1)}( \omega) + \delta_{16}F_1^{(1)}(
 \omega) - \delta_{a6}F_a^{(1)}( \omega) \Bigr] - \non
 && \hspace{-2ex} - \frac{4\varphi_s^{\eta} f_s}{v D_s} \beta \Bigl[ - \psi_6F^{(1)}(
 \omega) + \delta_{s6}F_s^{(1)}( \omega) + \delta_{16}F_1^{(1)}(
 \omega) - \delta_{a6}F_a^{(1)}( \omega) \Bigr] - \non
 && \hspace{-2ex} - \frac{2\beta}{v D_s}
 \Bigl[ \delta_{s6}^2 \cosh (2\tilde z + \beta \delta_{s6} \tilde
 \varepsilon_6)+ 4b\delta_{16}^2\cosh (\tilde z - \beta \delta_{16} \tilde
 \varepsilon_6) + \delta_{a6}^22a\cosh \beta \delta_{a6} \tilde
 \varepsilon_6^2 \Bigr], \nonumber
 \eea
The expressions for $F^{(1)}( \omega)$, $F_s^{(1)}( \omega)$, $F_a^{(1)}( \omega)$, $F_1^{(1)}( \omega)$ are given in \cite{0608U}.

 The longitudinal dynamic susceptibility of mechanically clamped ADP crystals, the piezoelectric coefficient, and
the elastic constant
are
 \bea && \hspace{-6ex}
  \chi_{33}^{\varepsilon}(\omega) = \chi_{33}^{\varepsilon 0} + \frac{\beta \mu_3^2}{v}F^{(1)}(\omega), \label{x33ea} \\
&& \hspace{-6ex}  e_{36}(\omega) = e_{36}^0 \!+\! \frac{\beta \mu_3}{v}
 \Bigl[ - 2\psi_6 F^{(1)}(\omega) \!+\! \delta_{s6} F_s^{(1)}(\omega)\! -\! \delta_{a6} F_a^{(1)}(\omega) \!+\! \delta_{16} F_1^{(1)}(\omega) \Bigr],  \label{e36a} \\
 && \hspace{-6ex} c_{66}^E(\omega) = c_{66}^{E0} + \label{c66Ea}\\
 && + \frac{4\beta \psi_6}{v D_a} \Bigl[ - 2\psi_6 F^{(1)}(\omega) + \delta_{s6} F_s^{(1)}(\omega)  + \delta_{16} F_1^{(1)}(\omega) - \delta_{a6} F_a^{(1)}(\omega) \Bigr] -  \nonumber\\
 && - \frac{4\varphi_a^{\eta} f_a}{v D_a} \beta
 \Bigl[ - 2\psi_6 F^{(1)}(\omega) + \delta_{s6} F_s^{(1)}(\omega)  + \delta_{16} F_1^{(1)}(\omega) - \delta_{a6} F_a^{(1)}(\omega) \Bigr] + \nonumber\\
 && + \frac{4\beta \psi_6}{v D_a} f_a - \frac{2\beta}{v D_a}
 \Bigl[ \delta_{s6}^2 a  + \delta_{16}^2 4b + \delta_{a6}^2 (1 + \cosh 2x)\Bigr], \nonumber
 \eea
The expressions for $F^{(1)}( \omega)$, $F_s^{(1)}( \omega)$, $F_a^{(1)}( \omega)$, $F_1^{(1)}( \omega)$ are given in \cite{0819U}.

In the static limit $\omega\rightarrow 0$ from (\ref{x33e})-(\ref{c66Ea}) we get
the isothermal static dielectric susceptibilities of a clamped crystal
\bea && \hspace{-8ex} \chi^{\varepsilon}_{33s,a} =  \chi^0_{33s,a} + {\bar
v}\frac{\mu^2}{v^2}\frac1T \frac{2\varkappa_{s,a}}{D_{s,a}-
2\varkappa_{s,a}\varphi^{\eta}_{s,a}}, \eea
 where \bea && \varkappa_s = \cosh
(2z_6 +\beta\delta_{s6}\varepsilon_6)
 + b\cosh (z_6 -\beta\delta_{16}\varepsilon_6) - (\eta_s^{(1)})^2D_s, ~~ \varkappa_a=a+b\cosh x, \nonumber \\
 &&  \varphi^\eta_s = \frac{1}{1- (\eta_s^{(1)})^2} +
\beta\nu_c, ~~ \varphi_c^{\eta} = \frac{1}{1-\eta_a^{(1)2}} + \beta \nu_c (0). \nonumber \eea
isothermal piezoelectric coefficients
\bea && e_{36s} =
e^0_{36s} + \frac{2\mu_3}{v}
\frac{\beta\theta_s}{D_s-2\varphi_s^\eta\varkappa_s}, \label{e36st} \\
&& e_{36a} = e_{36a}^0 + 2\frac{\mu_3}{v} \beta \frac{-2\varkappa_a +
f_a} {D_a - 2\varkappa_a \varphi_a^{\eta}}, \label{e36ast} \eea
where
\bea && \theta_s = - 2 \varkappa_s^c\psi_6 + f_s,~~~~
f_s=\delta_{s6} \cosh (2z_6 +\beta\delta_{s6}\varepsilon_6)
- \nonumber \\
&& -2b\delta_{16}\cosh(z_6
-\beta\delta_{16}\varepsilon_6)+\eta^{(1)z}(6)
(-\delta_{s6}M_{s6}+\delta_{a6}M_{a6}+\delta_{16}M_{16});\nonumber
\\
&& M_{a6}=2a\sinh \beta\delta_{a6}\varepsilon_6, ~~M_{s6}=\sinh
(2z_6 +\beta\delta_{s6}\varepsilon_6), ~~ M_{16}=4b \sinh (z_6
-\beta\delta_{16}\varepsilon_6). \nonumber
\\
 && f_a = \delta_{s6}a -
\delta_{16}2b\cosh x;  \nonumber \eea
isothermal elastic constants at constant field
 \bea
 && \hspace{-4ex} c_{66s}^{E} = c_{66s}^{E0} + \frac{8\psi_6}{v} \cdot
 \frac{\beta(- \psi_s \varkappa_s^c + f_s)}{D_s - 2\varphi_s^{\eta}
 \varkappa_s} -
 \frac{4 \beta\varphi_s^{\eta} f_s^2}{v D_s (D_s - 2\varphi_s^{\eta}
 \varkappa_s)} - \label{c66Est} \\
 && - \frac{2\beta}{v D_s} [\delta_{s6}^2 \cosh (2z_6 + \beta \delta_{s6} \varepsilon_6) +
 \delta_{a6}^2 2a\cosh \beta\delta_{a6}\varepsilon_6 + \nonumber\\
 && + \delta_{16}^2 4b \cosh (z_6 - \beta \delta_{16} \varepsilon_6)
 ] + \frac{2\beta}{v D_s^2}
 (-\delta_{s6}M_{s6}+\delta_{a6}M_{a6}+\delta_{16}M_{16})^2.  \nonumber \\
&& \hspace{-4ex} c_{66a}^E = c_{66a}^{E0} + \frac{8\psi_6}{v}
 \frac{\beta(-\psi_6 \varkappa_a + f_a)}{D_a - 2\varkappa_6 \varphi_a^{\eta}} -  \frac{4\beta \varphi_a^{\eta} f_a^2}{vD_a(D_a-2
\varkappa_a \varphi_a^{\eta})} - \label{c66East}\\
&& ~~~~- \frac{2\beta}{{v} D_a} (\delta_{16}^2 4b' \cosh x +
\delta_{s6}^2 a' + \delta_{a6}^2 2\cosh^2x).
 \nonumber \eea

Using the known relations between the elastic, dielectric, and piezoelectric characteristics we find

the isothermal elastic constants at constant polarization $ c_{66}^P =
c_{66}^E + e_{36}^2/\chi_{33}^{\varepsilon} $;
isothermal piezoelectric coefficients $d_{36} = \frac{e_{36}}{c_{66}^E}$;
isothermal dielectric susceptibilities at $\sigma = const$ $\chi_{33}^{\sigma} = \chi_{33}^{\varepsilon} + e_{36}d_{36}$.

\section{Experimental measurements of the thermodynamic characteristics}
\setcounter{equation}{0}
\renewcommand{\theequation}{6.\arabic{equation}}

To find the elastic constants and piezoelectric coefficients one should induce crystal vibrations of as simple form as possible and measure the resonant frequencies
($f_{r}$) of a metallized $45^{\circ}$ Z-cut plate, the antiresonant frequencies ($f_{a}$) of a non-metallized plate, as well as the crystal
capacity at low frequencies. The coefficient of electromechanical coupling represent the part of the electric energy transferred to the crystal
at zero frequency, which is transformed to the mechanical energy, and is determined by the frequencies $f_{r}$ and $f_{a}$:
 \bea
&&k^{2}_{36}=\frac{f_{a}^{2}-f_{r}^{2}}{f_{r}^{2}}.
\eea
The resonant frequency $f_{r}$ is given by the expression \cite{138x}:
 \bea
&&f_{r}=\frac{1}{2l}\sqrt{\frac{1}{\rho}s_{22}^{'E}},
\eea
where $\rho$ is the crystal density, and
 \bea
&&s_{22}^{'E}=\frac{1}{4}(s_{66}^{E}+s_{11}+s_{22}+2s_{12}),
\eea
where $s_{11}+s_{33}+2s_{13}$ are the elastic compliances, calculated in  \cite{138x}.
Using expressions (5.2) and  (5.3) we find the elastic compliance at constant field $s_{66}^{E}$
 \bea
&&s_{66}^{E}=\frac{1}{\rho l^{2}f_{r}^{2}}-(s_{11}+s_{22}+2s_{12}).
\eea
Respectively, the elastic constant at constant field is
 \bea
&&c_{66}^{E}=\frac{1}{\frac{1}{\rho l^{2}f_{r}^{2}}-(s_{11}+s_{22}+2s_{12})}.
\eea
The compliance at constant polarization is
 \bea
&&s_{66}^{P}=s_{66}^{E}(1-k_{36}^{2}).
\eea

The first resonant frequency of induced by the external a.c. field $ E_{3t}$=$E_{3}e^{i\omega t}$
vibrations of a thin square plate of a crystal with the sides $l$ cut in the  $(001)$ plane
and fixed along its perimeter,
reads \cite{cond2012}
 \bea
&&f_{r}=\frac{1}{l}\sqrt{\frac{c_{66}^{E}}{2\rho}}.
\eea
From where the elastic constant is
\bea
&&c_{66}^{E}= 2\rho l^{2}f_{r}^{2}.
\eea

Zwicker used another method to measure the elastic constants (see \cite{139x} ), by exciting the crystal vibrations with an external source
and observing the light diffraction by the ultrasound waves (Bergmann-Schaefer method). At this  $s^{E}$ and
$c^{E}$ are measured, even though the crystal is open-circuited.

The coefficient of piezoelectric strain $d_{36}$ is found from the following relation \cite{138x,371x}:
 \bea
&&d_{36}= 2k_{36}(\frac{\varepsilon_{33}^{\sigma}}{4\pi}s_{22}^{'E})^{\frac{1}{2}},
\eea
There are other ways to measure the coefficient  $d_{36}$ (see \cite{139x}). If the higher order
 effects are small, the following relation holds
 \bea
&&P_{3}= d_{36}\sigma_{6}~~~~~ (E=0).
\eea
Using the ballistic galvanometer  (E=0), Bantle and Caflish measured $P_{3}$ in KDP in the paraelectric phase
as a function of the stress $\sigma_{6}$ at different temperatures.

The converse piezoelectric effect is determined by the relation
\bea
&&\varepsilon_{6}= d_{36}E_{3}~~~~~ (\sigma_{6}=0).
\eea

Arx and Bantle applied the electric field $E_{3}$ and measured the elongation $\varepsilon_{1}'$ along [110]
\bea
&&\varepsilon_{1}'=\frac{1}{2}\varepsilon_{6}= \frac{1}{2}d_{36}E_{3}.
\eea
The obtained results are in a good agreement with the measurements in the direct piezoelectric effect.

The dielectric susceptibility of a mechanically free crystal is determined from the
measured capacity at low frequency
\bea
&&\varepsilon_{33}^{\sigma}=\frac{0.113 h}{S_{e}}C,
\eea
where $h$ is the distance between the electrodes; $S_{e}$ is the electrode area.
The dielectric permittivity of mechanically clamped crystal is calculated as
\bea
&&\varepsilon_{33}^{\varepsilon}=\varepsilon_{33}^{\sigma}(1-k_{36}^{2}).
\eea

\section{Comparison of the numerical results to experimental data. Discussion.}

\setcounter{equation}{0}
\renewcommand{\theequation}{7.\arabic{equation}}

Let us analyze the results of the numerical calculations of the dielectric, piezoelectric, elastic characteristics
of the KDP and ADP crystals nd compare them with
the corresponding experimental data.
It should be noted that the developed in the previous sections theory, strictly speaking, is valid for the
DKDP and DADP crystals only.
In view of the suppression of tunneling by the short-range
interactions \cite{48pok,135x,137x}, we shall assume that the presented in the previous sections results are valid for KDP and ADP as well.

For these calculations we use the values of the model parameters, which were found in  \cite{0608U,0819U} by fitting the theory to the experimental
temperature dependences of the physical characteristics of  KDP and ADP.  The used optimum set of the model parameters is given in Table~2.

\renewcommand{\arraystretch}{1.0}
\renewcommand{\tabcolsep}{3.0pt}
\begin{table}[!h]
\caption{The used set of the model parameters for KDP.}\label{tab3}
\begin{center}
\begin{tabular}{|c|c|c|c|c|c|c|c|c|c|}
\hline  & $T_c$ & $T_0$ & $\frac{\varepsilon}{k_B}$ & $\frac{w}{k_B}$ & $\frac{\nu_3(0)}{k_B}$ & $\mu_{3-}, 10^{-18}$ & $\mu_{3+}, 10^{-18}$ & $\chi_{33}^0$ \\
 & (K) & (K) & (K) & (K) & (K) & (esu$\cdot$cm) & (esu$\cdot$cm) & \\
\hline   KDP & 122.5 & 122.5 &  56.00 &  422.0 &  17.91 &
1.46 &   1.71 &   0.73 \\ \hline
\end{tabular}
%\end{center}
%\begin{center}
\begin{tabular}{|c|c|c|c|c|c|c|c|c|c|}
\hline  & $\frac{\psi_{6}}{k_B}$ & $\frac{\delta_{s6}}{k_B}$ & $\frac{\delta_{a6}}{k_B}$ & $\frac{\delta_{16}}{k_B}$ & $c_{66}^0\cdot 10^{-10}$ & $e_{36}^0$ \\
 & ($K$) & ($K$) & ($K$) & ($K$) & (dyn/cm$^2$) & (esu/cm) \\
\hline   KDP &-150.00 &  82.00 &-500.00 &-400.00  &   7.10
&1000.00 \\
 \hline
\end{tabular}
\end{center}
%%\begin{center}
%\begin{tabular}{|c|c|c|c|c|c|c|c|c|c|}
%\hline $x$ & $P_-$ & $R_-$ & $P_+$ & $R_+$ \\
% & ($s$) & ($\frac{s}{K}$) & ($s$) & ($\frac{s}{K}$) \\
%\hline     0.00 &   0.35 & 0.0100 &   0.43 & 0.0160 \\
% \hline
%\end{tabular}
\end{table}
\renewcommand{\arraystretch}{1}
\renewcommand{\tabcolsep}{1pt}

\renewcommand{\arraystretch}{1.0}
\renewcommand{\tabcolsep}{3.0pt}
\begin{table}[!h]
\caption{The used set of the model parameters for ADP.}\label{tab31}
\begin{center}
\begin{tabular}{|c|c|c|c|c|c|c|c|c|}
\hline
  & $T_N,$ & $\frac{\varepsilon^{'0}}{k_B},$ & $\frac{w^{'0}}{k_B},$ & $\frac{\nu_c^{0}}{k_B},$ & $\mu_3, 10^{-18}$, & $\chi_{33}^{0\varepsilon}$ \\
  & (K) & (K) & (K) & (K) & (esu$\cdot$cm) &   \\
\hline
ADP & 148 & 20 & 490,0 & -10,00 & 2,10 & 0,23 \\
\hline
\end{tabular}
\begin{tabular}{|c|c|c|c|c|c|c|c|c|}
\hline
  & $\frac{\psi_6}{k_B}$, & $\frac{\delta_{s6}}{k_B}$, & $\frac{\delta_{a6}}{k_B}$, & $\frac{\delta_{16}}{k_B}$, & $c_{66}^0\cdot 10^{-10}$ & $e_{36}^0$ \\
  & (K) & (K) & (K) & (K) & (dyn/cm$^2$) & (esu/cm) \\
\hline
ADP & -160 & 1400 & 100 & -300 & 7.9 & 10000 \\
\hline
\end{tabular}

%\begin{tabular}{|c|c|c|c|c|c|c|c|c|}
%\hline
%  & $\frac{\delta_{s1}}{k_B}$, & $\frac{\delta_{s2}}{k_B}$, & $\frac{\delta_{s3}}{k_B}$, & $c_{11}^0\cdot 10^{-10}$, & $c_{12}^0\cdot 10^{-10}$ & $c_{13}^0\cdot 10^{-10}$ & $c_{33}^0\cdot 10^{-10}$ \\
%  & (K) & (K) & (K) & ($dyn/cm^2$) & ($dyn/cm^2$) & ($dyn/cm^2$) & ($dyn/cm^2$) \\
%\hline
%ADP & 100 & 100 & 100 & 78 & -22 & 14.5 & 30.5 \\
%\hline
%\end{tabular}
\end{center}
\end{table}
\renewcommand{\arraystretch}{1}
\renewcommand{\tabcolsep}{1pt}

The energy $w_{1}$ of two proton configurations with four or zero
protons near the given oxygen tetrahedron should be much higher than
$\varepsilon$ and $w$. Therefore we take $w_{1H} = \infty$ and
$w_{1}=\infty$ $(d=0)$ and $w'_{1}=\infty$ $(d'=0)$.

The primitive cell volume, containing two PO$_{4}$ groups is taken to be equal
$v=0.1936\cdot 10^{-21}$ cm$^{3}$ for KDP and
 $v=0.2110\cdot 10^{-21}$ cm$^3$  for ADP.

The calculated temperature dependences of  static dielectric
permittivities of a free $\varepsilon
_{33}^\sigma $
\begin{figure}[!h]
\begin{center}
 \includegraphics[scale=0.5]{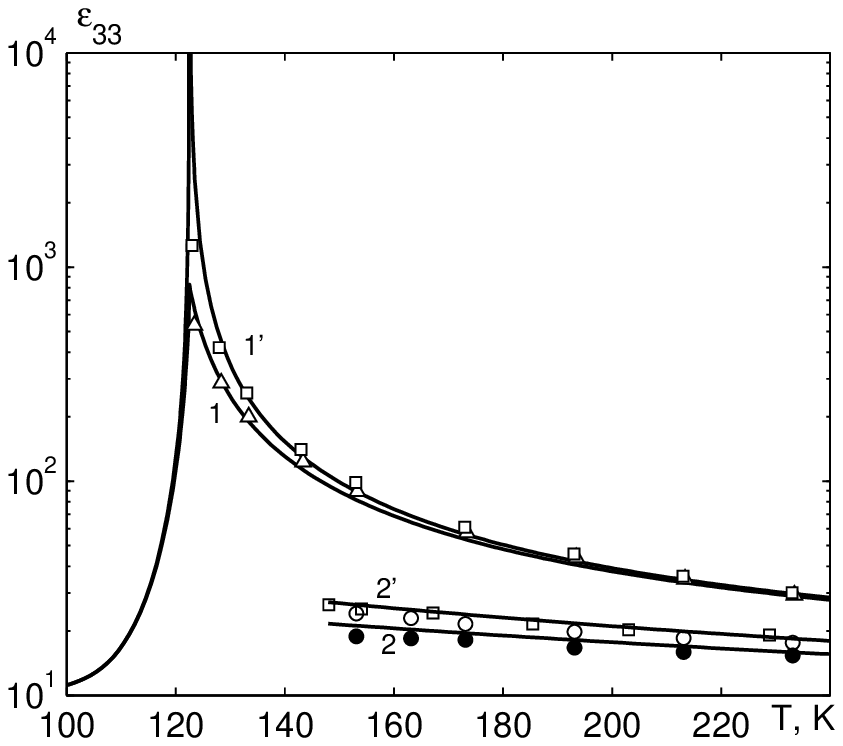}~~~~\includegraphics[scale=0.5]{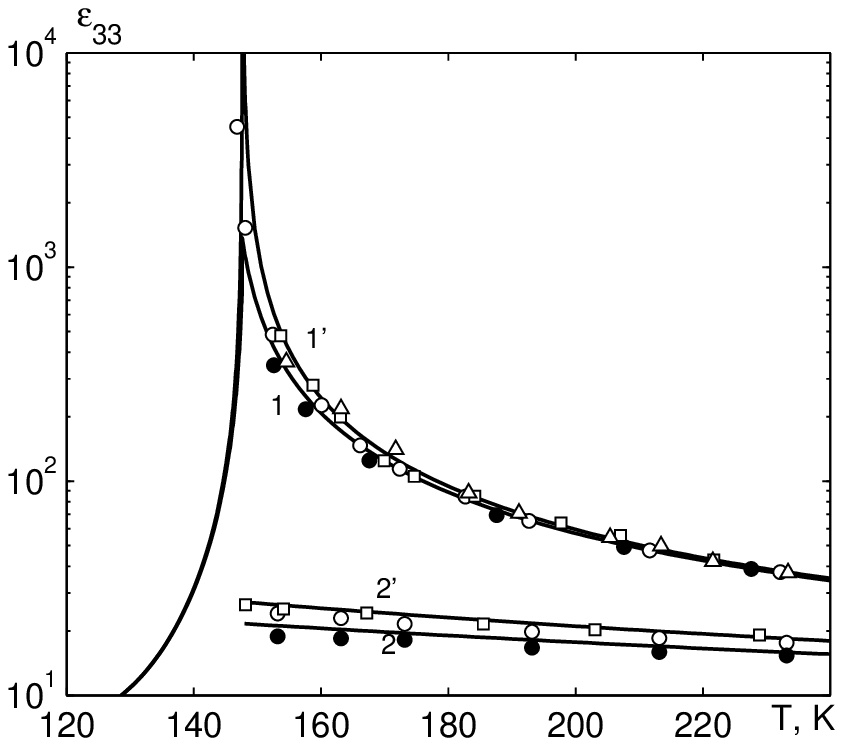}
\end{center}
\caption[]{Temperature dependences of the dielectric
permittivity of a clamped
($\varepsilon _{33}^\varepsilon )$ (1) and free $\varepsilon _{33}^\sigma $ (1')
 KDP crystals: \ra{\e{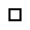}},
\ra{\e{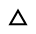}}\cite{138x}; free (2) and clamped (2')
ADP crystals: $\bullet$ \cite{138x},   $\circ$ \cite{138x}, \ra{\e{s0.eps}} \cite{485x}.} \label{e33_KDP_ADP}
\caption[]{Temperature dependences of the dielectric
permittivity of a clamped
($\varepsilon _{33}^\varepsilon )$ (1) and free $\varepsilon _{33}^\sigma $ (1')
RDP: $\circ$ %\ra{\e{o0.eps}}
\cite{371x},
$\bullet$ %\ra{\e{o1.eps}}
(($\varepsilon^{\sigma}_{33}$-1)/4$\pi$\cite{367x}-
$d_{36}^{2}$\cite{367x}/$s_{66}^{E}$)4$\pi$+1,$\Box$
%\ra{\e{s0.eps}}
\cite{403x}, $\bigtriangleup$
%\ra{\e{up0.eps}}
\cite{379x}; free (2) and clamped (2')
 ADP: $\bullet$ \cite{138x},   $\circ$ \cite{138x}, \ra{\e{s0.eps}} \cite{485x}.} \label{e33_RDP_ADP}
\end{figure}

and clamped $\varepsilon _{33}^\varepsilon $
KDP, RDP, and ADP crystals along with the experimental data shown in figs.~\ref{e33_KDP_ADP}-\ref{e33_RDP_ADP}.
At approaching $T_{c }$  in the paraelectric phase  $\varepsilon
_{33}^{\sigma} $ increases by the hyperbolic law at approaching the phase transition, reaching very high values at $T = T_{c}$.
Below the phase transition $\varepsilon _{33}^{\sigma} $
decreases rapidly.
The temperature behavior of ($\varepsilon _{33}^{\sigma} )^{-1}$
obeys the Curie-Weiss law is obeyed in the temperature
range  $\Delta $T $<$ 50 K, and an essential non-linearity of the temperature curve of
($\varepsilon _{33}^{\sigma} )^{-1}$ is observed. The calculated without taking
into account the piezoelectric coupling dielectric permittivity
$\varepsilon_{33 }$ of KDP at $\Delta $T $<$ 50 K coincides with $\varepsilon
_{33}^{\sigma} $, whereas at larger $\Delta $T  the curve of $\varepsilon_{33}(T) $ is lower than that of  $\varepsilon
_{33}^{\sigma} (T)$. The permittivity  $\varepsilon^{\sigma}_{33}$ of ADP is by $\sim18\%$
larger than $\varepsilon^{\varepsilon}_{33}$; the difference between the two characteristics is temperature independent.

\begin{figure}[!h]
\begin{center}
 \includegraphics[scale=0.5]{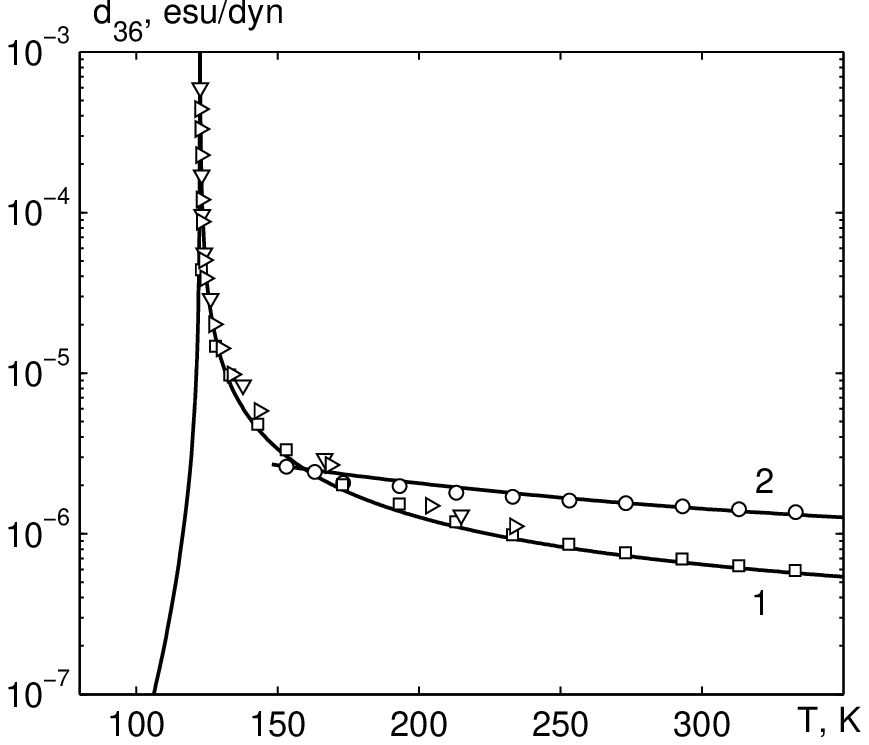}~~~~\includegraphics[scale=0.5]{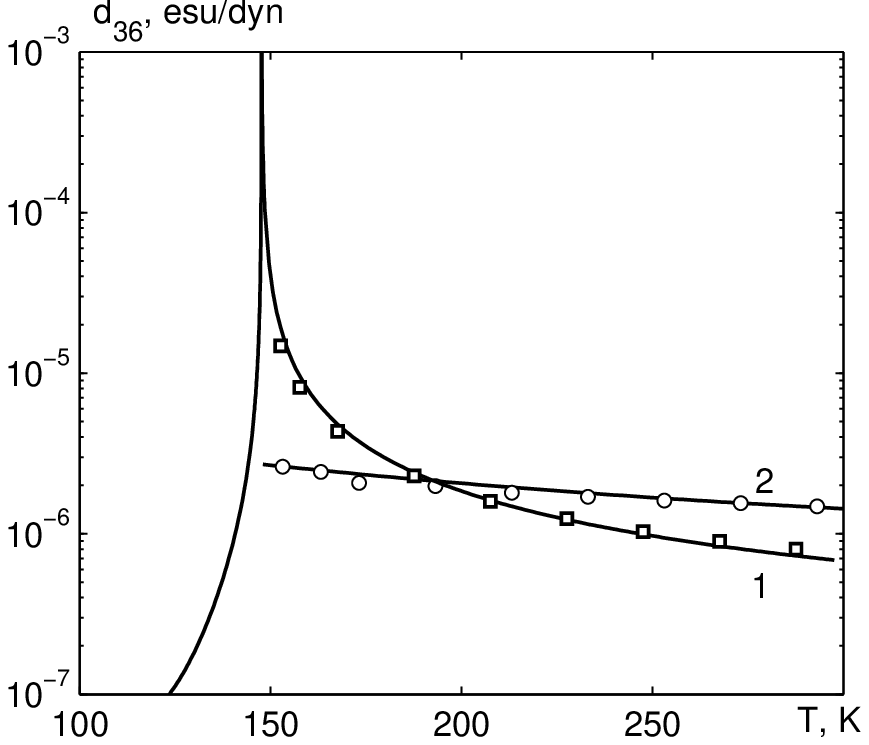}
\end{center}
\caption[]{Temperature dependence of  the
coefficients of the piezoelectric strain $d_{36}$ of KDP -- 1,
\ra{\e{s0.eps}}\cite{138x}, \ra{\e{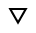}}\cite{bantle43},
\ra{\e{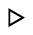}}\cite{von43};
ADP -- 2, $\circ$,
\cite{138x}. Lines: the theory.} \label{d36_KDP_ADP}
\caption[]{Temperature dependence of  the
coefficients of the piezoelectric strain $d_{36}$ of  RDP  -- 1,
\ra{\e{s0.eps}}\cite{371x};
ADP -- 2, $\circ$,
\cite{138x}. Lines: the theory.} \label{d36_RDP_ADP}
\end{figure}

In figs.~\ref{d36_KDP_ADP}-\ref{d36_RDP_ADP} we show the
calculated temperature dependences of  the coefficients of the
piezoelectric strain $d_{36}$ of  KDP, RDP, and ADP.

\begin{figure}[!h]
\begin{center}
  \includegraphics[scale=0.5]{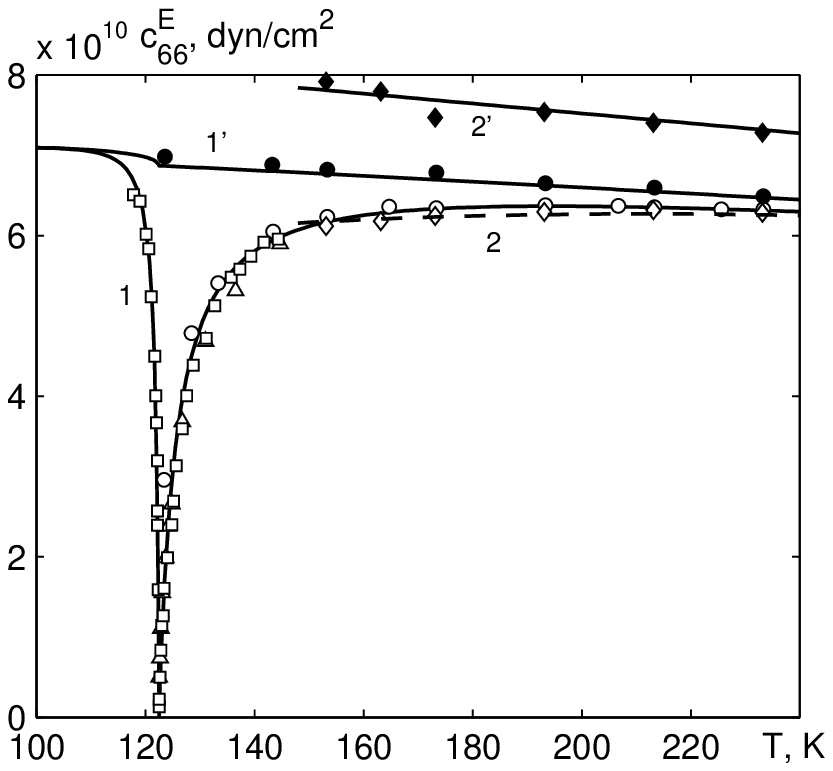}~~~ \includegraphics[scale=0.5]{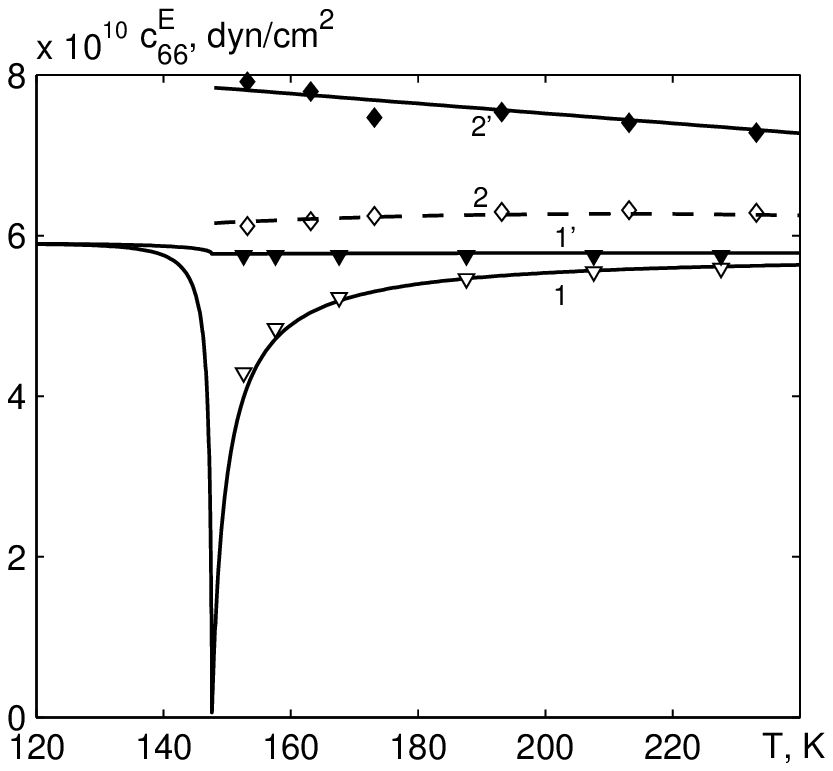}
\end{center}
\caption[]{Temperature dependences of the elastic constants $c_{66}^E $ and $c_{66}^P $ of
KDP (1, 1', respectively): \ra{\e{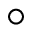}}\cite{138x},
\ra{\e{s0.eps}}\cite{brody68}, \ra{\e{up0.eps}}\cite{385x}; and ADP (2, 2'):
\ra{\e{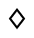}}, \cite{138x}.}
\label{c66_KDP_ADP}
\caption[]{Temperature dependences of the elastic constants $c_{66}^E $ and $c_{66}^P $ of
RDP (1, 1', respectively): \ra{\e{do0.eps}}--$1/s_{66}^{E}$\cite{shuv66}; and ADP (2, 2'):
\ra{\e{d0.eps}}, \cite{138x}.}
\label{c66_RDP_ADP}
\end{figure}

The temperature dependences of $c_{66}^{E}$ and $c_{66}^{P}$ of
KDP, RDP and ADP are shown in fig.~\ref{c66_KDP_ADP} and \ref{c66_RDP_ADP}.
At the transition temperature the elastic constant $c_{66}^E $ of
 KDP  approaches zero. The temperature dependence of  $c_{66}^P $ has no anomaly at the phase transition.
The elastic constant  $c_{66}^E$ of ADP, in contrast to KDP, is finite at $T=T_N$ and hardly temperature dependent.

\begin{figure}[!h]
\begin{center}
 \includegraphics[scale=0.7]{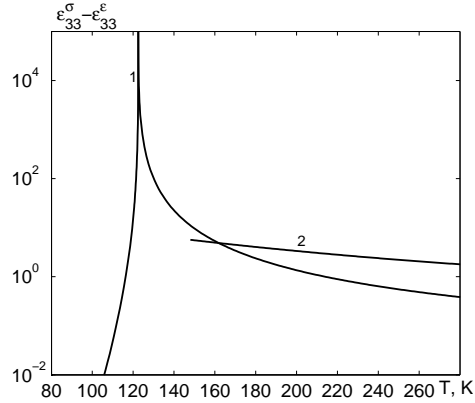}
\end{center}
\caption[]{The temperature dependence of the difference $\varepsilon_{33}^{\sigma}-\varepsilon_{33}^{\varepsilon}$
of KDP (1) and ADP (2).} \label{e33s_e33e_KDP_ADP}
\end{figure}

The temperature dependence of the difference
 $\varepsilon_{33}^{\sigma}-\varepsilon_{33}^{\varepsilon}=4\pi e_{36}d_{36}=4\pi e_{36}^2/c_{66}^E=4\pi d_{36}^2 c_{66}^E$
in KDP and ADP is plotted in fig.~ \ref{e33s_e33e_KDP_ADP}

Hence, the proposed theory, as seen in figs.~\ref{e33_KDP_ADP}-\ref{c66_KDP_ADP} adequately describes
the experimental data for the static dielectric, piezoelectric, and elastic characteristics of  KDP and ADP.

Let us calculate the longitudinal dynamic characteristics of mechanically free KDP and ADP crystals, cut as  $l\times l$  square plates
 ($l=1$~mm) in the (0,0,1) plane.

\begin{figure}[!h]
\begin{center}
 \includegraphics[scale=0.7]{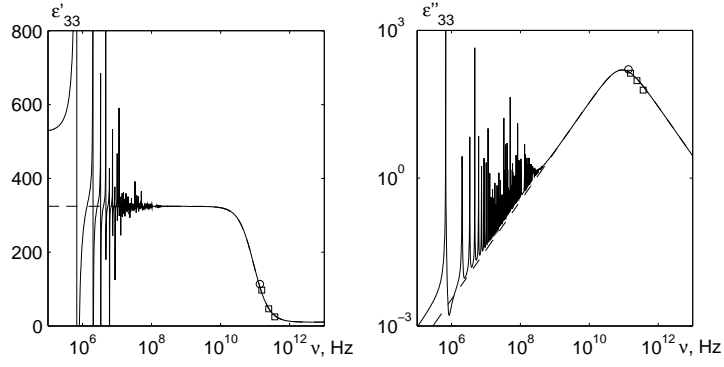}
\end{center}
\caption[]{The frequency dependences of the real and imaginary parts of the dielectric susceptibility of
free and clamped
 KDP at $\Delta T$=5K,  $\circ$ -- \cite{398x},
$\square$ -- \cite{408x}.} \label{e33es_nuKDPdT5}
\end{figure}

In figs.\ref{e33es_nuKDPdT5} and \ref{e33snu} we show the
frequency dependences
 of the real and imaginary parts of the dielectric susceptibility of
free
 KDP  at $\Delta T$=5K, RDP at $\Delta T$=5, 10, 50~K, and ADP at $\Delta T$=28~K. In the frequency range $3\cdot 10^5$--$3\cdot 10^8$~Hz the susceptibility
 of these crystals has a resonant dispersion. The resonant frequencies are inversely proportional to
 the sample side length. The dashed lines in figs.~\ref{e33es_nuKDPdT5}, \ref{e33snu} correspond to the low-frequency
 permittivity curve of a clamped crystal. With increasing frequency or temperature $\Delta T$ the resonant peaks lower down. The last
 peak shifts to higher frequencies with increasing temperature $\Delta T$. A similar multi-peak resonant dispersion is also observed in the
 ferroelectric phase.
 Above the resonances the crystal is clamped by the high-frequency field; the permittivity of a clamped crystal above has a relaxational
 dispersion above $10^9$~Hz. At
 $\omega \to 0$ we obtain the static dielectric permittivity of  a free crystal.

\begin{figure}[!h]
\begin{center}
 \includegraphics[scale=0.5]{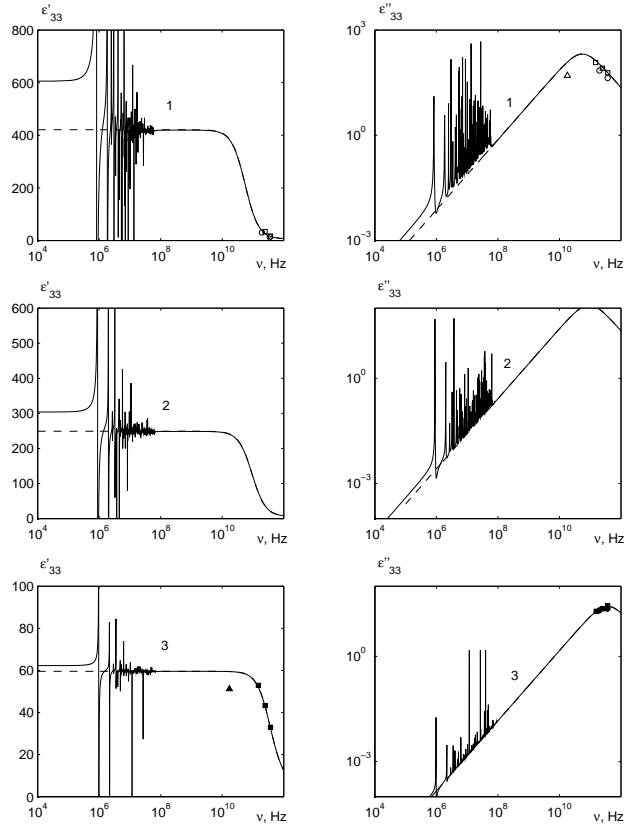}
\end{center}
\caption[]{The frequency dependences of the real and imaginary parts of the dielectric susceptibility of
free and clamped
 RDP at $\Delta T$=5K- 1, $\Delta T$=5K- 2,$\Delta T$=5K- 3; $\circ$ -- \cite{398x},
$\square$ -- \cite{408x}.} \label{e33es_nuRDPdT5_10_50_kl}
\end{figure}

\begin{figure}[!h]
\begin{center}
 \includegraphics[scale=0.65]{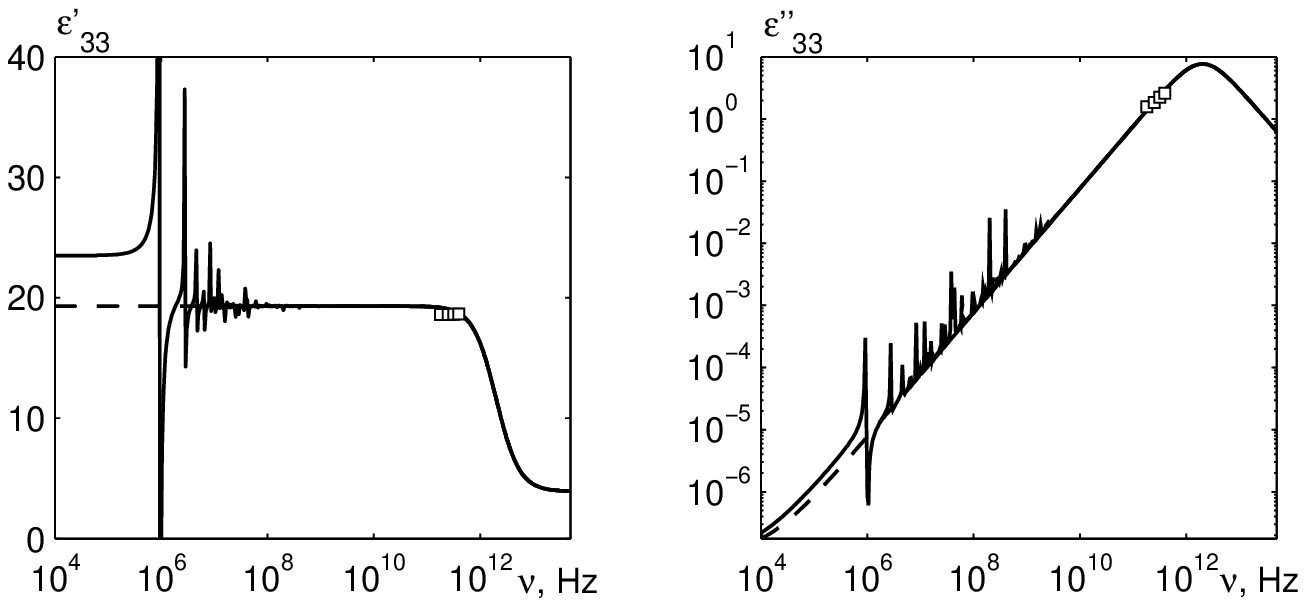}
\end{center}
\caption[]{The frequency dependences of the real and imaginary parts of the dielectric susceptibility of
free and clamped
 (dashed line)
 ADP at $\Delta T = 28$~K, $\square$ --  \cite{465x}.}
\label{e33snu}
\end{figure}

Below the piezoelectric resonances, where the free crystal permttivity is measured, $k_3^2(\nu)\approx0.4$,
whereas above the resonances, where the clamped crystal permittivity is measured, $k_3^2(\nu)\approx0$.

\begin{figure}[!h]
\begin{center}
 \includegraphics[scale=0.6]{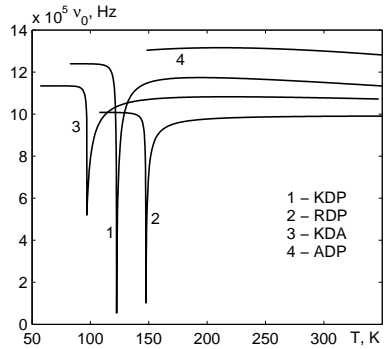}
\end{center}
\caption[]{The temperature dependences of the lowest resonance
frequencies (k = l = 0) of the KDP, RDP, KDA, and ADP crystals. }
\label{nu0_KDP_DADP}
\end{figure}

In fig.~\ref{nu0_KDP_DADP}
we plot the temperature dependences of the lowest resonance frequencies (k = l = 0) of the KDP, RDP, KDA, and ADP crystals
at $L_{x}=1,37$cm and $L_{y}=0,79$~cm.

The presented here results show that the presence of the piezoelectric coupling in the  KH$_{2}$PO$_{4}$ family crystals
leads to the difference between the permittivities of mechanically free and clamped crystals, existence of the piezoelectric coefficients, and piezoelectric
resonances. Unfortunately, such studies, either theoretical or experiment, have not been performed yet
for the mixed compounds of the  K$_{1-x}$(NH$_4$)$_x$H$_2$PO$_4$ type.
It is important to explore the temperature and frequency dependences
of ${\varepsilon }'_{33}(\nu T)$ and ${\varepsilon }''_{33}(\nu T)$, estimate the coefficient of
electromechanical coupling and the difference between the free and clamped permittivities, which indicates
the presence of the piezoeffect. It is also interesting to explore the concentrational dependences of the above mentioned characteristics in
the  K$_{1-x}$(NH$_4$)$_x$H$_2$PO$_4$ type systems.

\section{Dielectric, piezoelectric, and elastic characteristics of the ferroelectrics K$_{1-x}$(NH$_4$)$_x$H$_2$PO$_4$ systems.}

\begin{figure}[!h]
\begin{center}
 \includegraphics[scale=0.8]{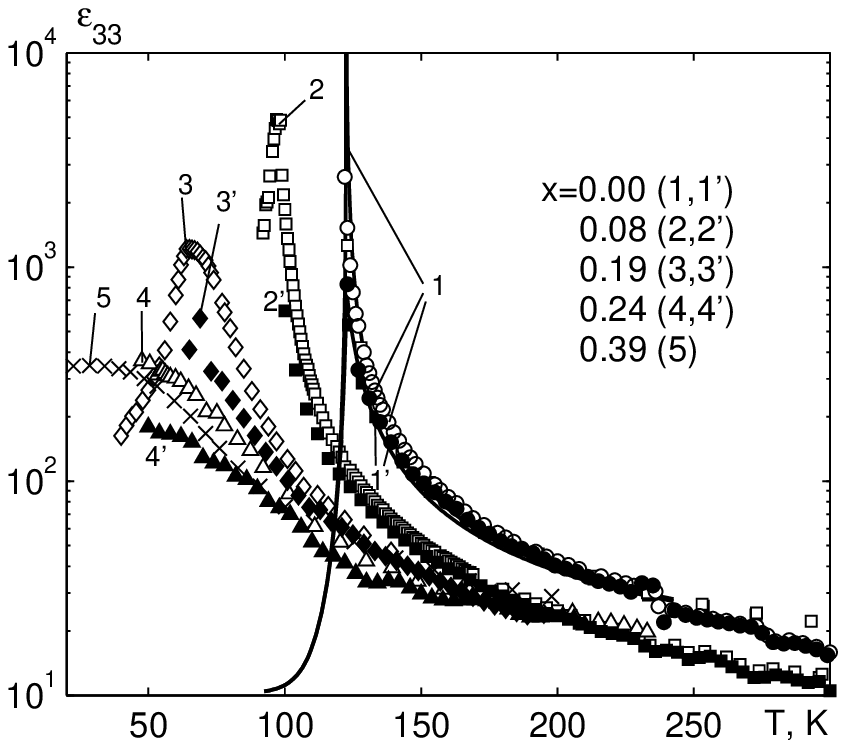} ~~~~\includegraphics[scale=0.8]{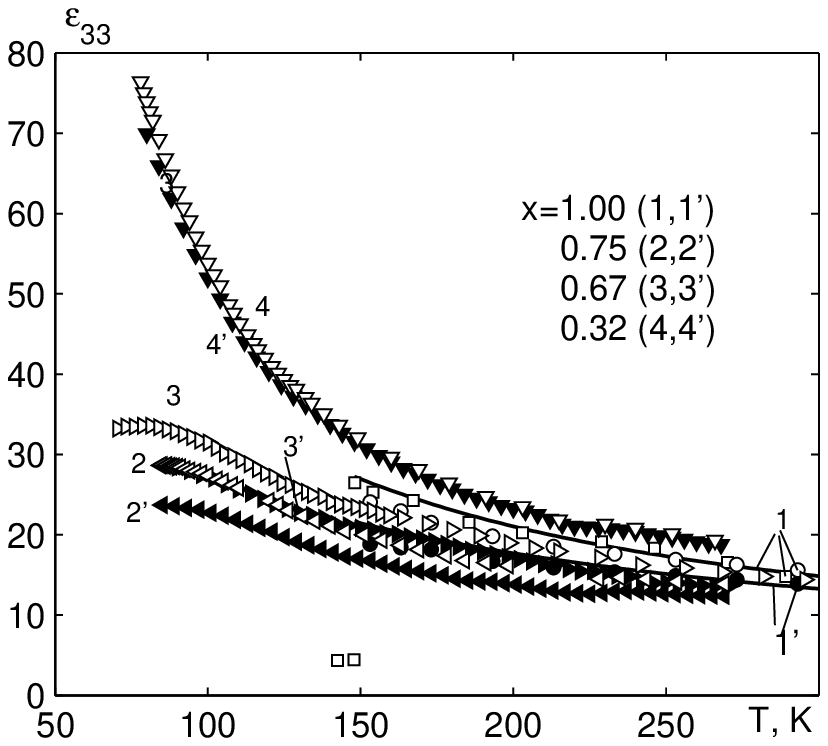}
 a~~~~~~~~~~~~~~~~~~~~~~~~~~~~~~~~~~~~ b
\end{center}
\caption[]{The temperature dependences of the dielectric permittivity of the mixed K$_{1-x}$(NH$_4$)$_x$H$_2$PO$_4$ crystals
at small $x$ (a) and large $x$ (b). The data marked as 1-4 and 1'-4' correspond to the
permittivities of fee and clamped samples, respectively. Solid lines are the theoretical results of  \cite{JPS1701}. Data represented by $\bullet$ and   $\circ$
are taken from  \cite{138x}.
 } \label{e33_KDP_Kor_prep}
\end{figure}

\begin{figure}[!h]
\begin{center}
 \includegraphics[scale=0.8]{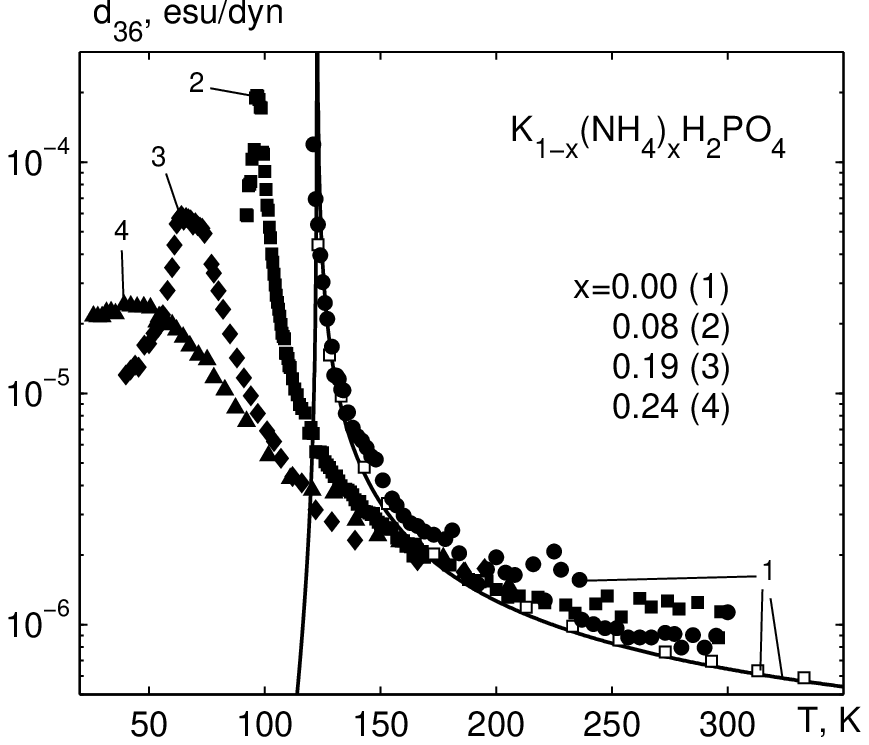} ~~~\includegraphics[scale=0.8]{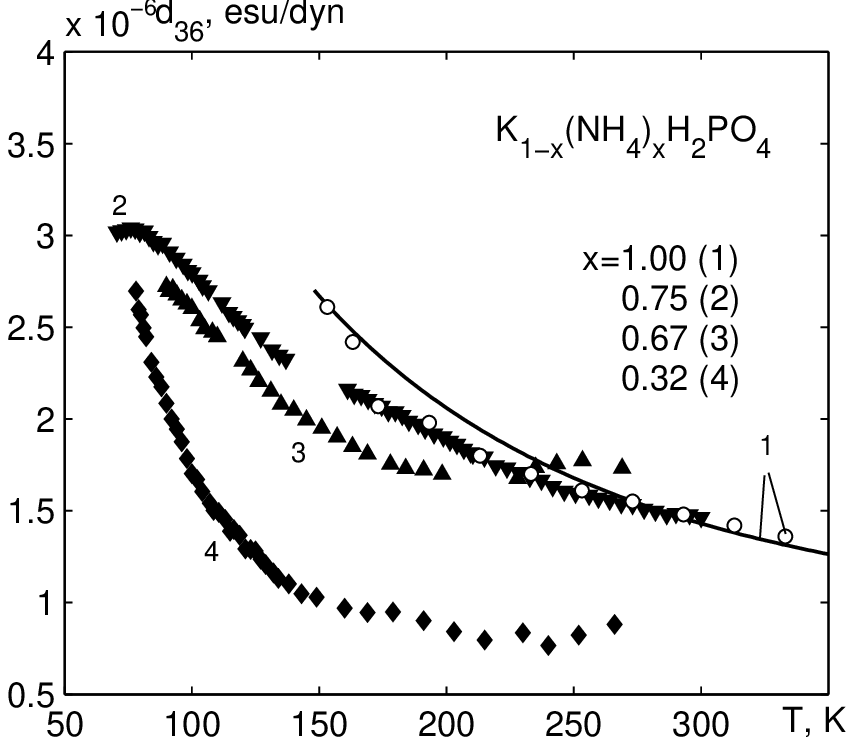}
           a~~~~~~~~~~~~~~~~~~~~~~~~~~~~~~~~~~~~ b
\end{center}
\caption[]{The temperature dependences of the coefficient of piezoelectric strain of  K$_{1-x}$(NH$_4$)$_x$H$_2$PO$_4$ at small $x$ (a) and large $x$ (b).
Solid lines are the theoretical results of  \cite{JPS1701}. Data represented by \ra{\e{s0.eps}} are taken from
\cite{138x}).
}
\label{d36_KDP_Kor_prep}
\end{figure}

Experimental measurements were performed for the samples with
ammonium content $x = 0.0$, 0.08;  0.19, 0.24, 0.32, 0.67, 0.75
and 0.97, having the form of thin plates cut at 45$^\circ$ to the
axes \textbf{a}, \textbf{b}, and perpendicularly to the axis
\textbf{c}. Silver electrodes were evaporated on the surfaces
perpendicular to the axis \textbf{c}. The samples were placed in
in a cryostat, where the temperature was varied between 10 and
300~K within an accuracy less than  0.5~K. The dielectric
permittivities ($\varepsilon_{33}^{\sigma}$ and
$\varepsilon_{11}^{\sigma}$) were measured using the capacitance
bridge operating at 10~kHz. The resonance $f_{r}$ and
antiresonance $f_{A}$ frequencies were measured. The elastic
constant $c_{66}^{E}$ was calculated using the relation (6.5). The
elastic compliances and densities, occurring in (6.5), for
K$_{1-x}$(NH$_4$)$_x$H$_2$PO$_4$ were calculated in the mean
crystal approximation
\[ \rho(x)=\rho^{KDP}(1-x)+\rho^{ADP}x; ~~~~S_{ij}(x) =S_{ij}^{KDP}(1-x)+S_{ij}^{ADP}x.
\]
The experimental values of $\rho^{KDP}$, $\rho^{ADP}$, $S_{ij}^{KDP}$, $S_{ij}^{ADP}$ were taken from  \cite{138x}.
The piezoelectric coefficients $d_{36}$ and dielectric permittivities $\varepsilon_{33}^{\varepsilon}$ of the K$_{1-x}$(NH$_4$)$_x$H$_2$PO$_4$
crystals are calculated using (6.9) and (6.14), respectively.
The experimental results are presented in figs.~\ref{e33_KDP_Kor_prep}, \ref{d36_KDP_Kor_prep}.

As one can see, $\varepsilon _{33}^{\sigma } $ and $d_{36} $ increase with decreasing temperature  and
have maxima at the ferroelectric phase transition temperatures   (\textit{Tc}), or the transition to the proton glass phase (\textit{Tg}). The shown in figs.~\ref{e33_KDP_Kor_prep},
\ref{d36_KDP_Kor_prep} dependences indicate the existence of piezoelectricitiy in the mixed K$_{1-x}$(NH$_4$)$_x$H$_2$PO$_4$ crystals
for all studied compositions $x$.
With increasing ammonium content, the maxima of the $\varepsilon _{33}^{\sigma }(T) $ and $d_{36} (T)$ shift to lower temperatures, get smeared out, and lower
down. Hence, at $x\approx 0.24$, undergoing the phase transition into the proton glass state the values of the piezoelectric coefficient an dielectric permittivity
at quite large at temperatures well below Tg, whereas in the ferroelectric compounds $\varepsilon _{33}^{\sigma } $ and $d_{36} $ decrease
very fast in the low-symmetry phase.

The temperature dependences of the elastic constant $c_{66}^{E} $ have minima in the vicinities of the
corresponding transition temperatures (fig.~11). %\label{c66_KDP_Kor_prep}).

Caused by the electromechanical coupling correlation in behavior of the  $c_{66}^{E}(T) $ and $\varepsilon _{33}^{\sigma }(T) $ curves is observed.
As follows from the form the  $d_{36}(T) $ dependence, this coupling is the largest at
temperatures, where the minimum of  $c_{66}^{E} $ is observed,
which coincides with the position of the dielectric permittivity maximum.

\begin{figure}[!h]
\begin{center}
 \includegraphics[scale=0.8]{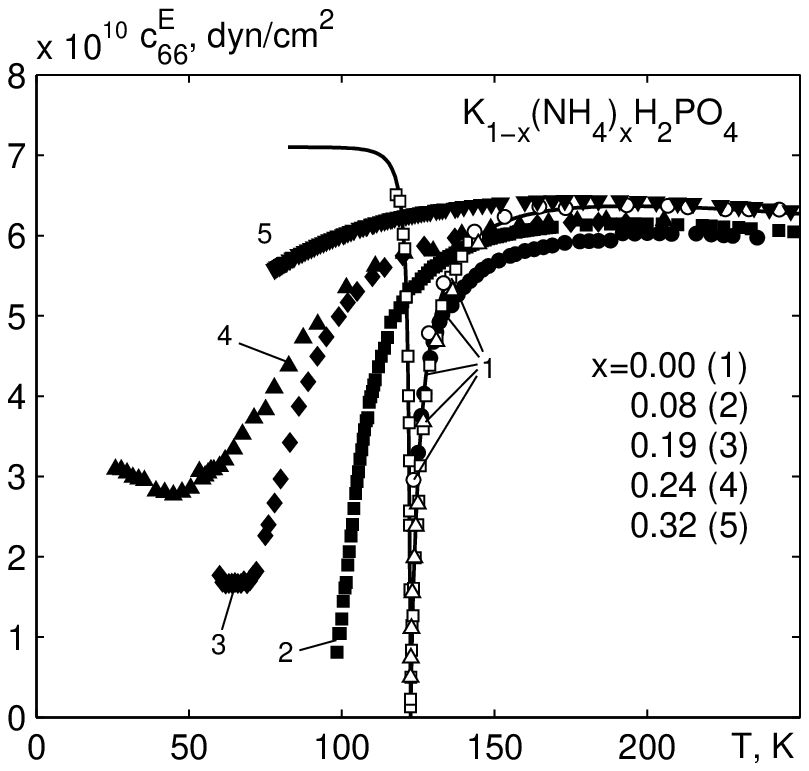}~~~~ \includegraphics[scale=0.8]{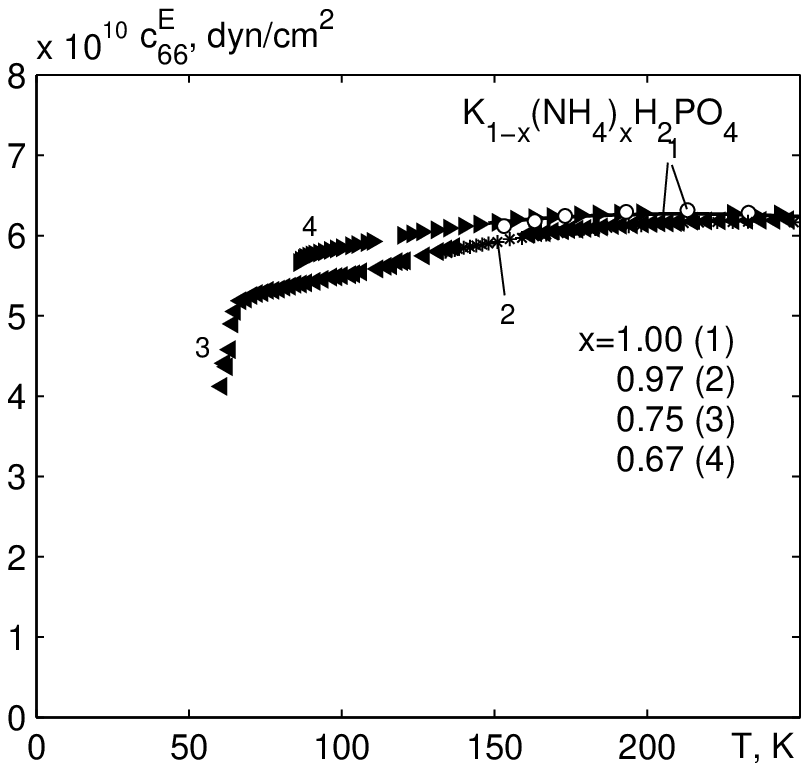}
           a~~~~~~~~~~~~~~~~~~~~~~~~~~~~~~~~~~~~ b
\end{center}
\caption[]{The temperature dependences of the of the elastic constant of K$_{1-x}$(NH$_4$)$_x$H$_2$PO$_4$
  at small $x$ (a) and large $x$ (b).
Solid lines are the theoretical results of  \cite{JPS1701}. Data
for $x=0$ are taken from \cite{138x} $\circ$, \cite{brody68}
$\square$, \cite{385x}  $\vartriangle$. } \label{c66_KDP_Kor_prep}
\end{figure}

\section{Theory of dielectric, piezoelectric, and elastic properties
of mixed  K$_{1-x}$(NH$_4$)$_x$H$_2$PO$_4$ crystals}

In the cluster approximation the mixed K$_{1-x}$(NH$_4$)$_x$H$_2$PO$_4$ crystal
can be presented as a system of independent PO$_4$ tetrahedra ($"A"$,$"B"$,$"A'"$,$"B'"$) in an effective
field with four effective dipole moments $\left\langle \vec{d}_{f} \right\rangle _{c} $ of the proton bonds, where
$\left\langle ...\right\rangle _{c} $ stands for averaging over the compositional configurations \cite{Vdovych523,cmp13706,Vdovych101,ferro43,izv1268}.
The contribution of heavy ions is taken into account via renormalization of
the Hamiltonian parameters. The effective pseudospin cluster Hamiltonian (e.g. of the  $"B"$ type) can be written as

\begin{equation} \label{GrindEQ__1_}
\begin{array}{l} {H_{\alpha }^{B} \left(\left\{\xi _{f} \right\};\left\{S_{f} \right\}\right)=H_{cl,\alpha }^{B} \left(\left\{S_{f} \right\}\right)+V_{\alpha }^{B} \left(\left\{S_{f} \right\}\right)\varepsilon _{6} -\sum _{f=1}^{4}\xi _{f} S_{f}  ;\quad } \\ {\xi _{f} =\left\langle \vec{d}_{f} \right\rangle _{c} \cdot \vec{E}+\varphi _{f} +\varphi _{L,f} +\left\langle d_{f}^{z} \right\rangle _{c} \cdot \left(\psi _{6,f} \cdot \varepsilon _{6} +G_{f}^{z} \right)} \end{array} .
\end{equation}

Here $\vec{E}$ is a uniform external electric field; $G_{f}^{z} $
is the z-component of the internal random deformational field;
$\varphi _{f} $  is the random cluster field, caused by the
influence of the  \textit{``A''} tetrahedron on the  bond
\textit{f}, connecting the \textit{``A''} and \textit{``B''}
tetrahedra; $\varphi _{L,f} $ is the random long-range field,
created by all effective dipoles apart from those belonging to the
\textit{``A''} and \textit{``B''} tetrahedra. The coefficient of
the piezoelectric coupling is presented as a product $\left\langle
d_{f}^{z} \right\rangle _{c} \psi _{6,f} $, where the variable
$\psi _{6,f} =\left\langle \psi _{6,f} \right\rangle +A_{\psi _{6}
} \cdot G_{f}^{z} $ describes its spatial fluctuations, whereas
$\varepsilon _{6} $ is the component of the strain tensor; the
pseudospin $S_{f} =\pm 1$ describes the proton position on the
hydrogen bonds.

Only two lowest energy levels $\varepsilon _{\alpha } $ and $w_{\alpha } $
 (Slater and Takagi configurations) of protons around the $PO_{{\rm 4}} $ group are taken into account for the
 Hamiltonian  $H_{cl,\alpha }^{B} \left(\left\{S_{f} \right\}\right)$; it is also assumed that the highest levels $\; w_{1\alpha } \to \infty $.
 The quantity $V_{\alpha }^{B} \left(\left\{S_{f} \right\}\right)$ describes the splitting of the lowest energy levels due to the strain  $\varepsilon _{6} $.
 we shall asume that an arbitrary cluster can be either in the configurational state $\alpha =+$ with the energy parameters $\varepsilon _{+} ,\; w_{+} $
with the probability $c_{+} =1-x$ or in the state $\alpha =-$ with the energy parameters $\varepsilon _{-} ,\; w_{-} $ with the probability
 $c_{-} =x$. Then the state of the effective dipole moment $\vec{d}_{f} (\alpha _{f} ,\alpha '_{f} )$ corresponding to the bond $f$ is determined by the
 states  $\alpha _{f} ,\alpha '_{f} $ of the two tetrahedra connected by this bond; the average bond moment $\left\langle \vec{d}_{f} \right\rangle _{c} $
 reads:

\begin{equation} \label{GrindEQ__2_}
\begin{array}{l} {\left\langle \vec{d}_{f} \right\rangle _{c} \approx c_{+}^{2} \vec{d}_{f+} +c_{-}^{2} \vec{d}_{f-} +2c_{+} c_{-} \vec{d}_{f0} ;\quad \vec{d}_{f\pm } =\vec{d}_{f} (\pm ,\pm );\; \; \vec{d}_{f0} =\vec{d}_{f} (+,-)=\vec{d}_{f} (-,+);} \\ {\vec{d}_{1\lambda } =(d_{\lambda }^{x} ,0,d_{\lambda }^{z} ),\; \vec{d}_{3\lambda } =(-d_{\lambda }^{x} ,0,d_{\lambda }^{z} ),\; \vec{d}_{2\lambda } =(0,-d_{\lambda }^{y} ,d_{\lambda }^{z} ),\; \vec{d}_{4\lambda } =(0,d_{\lambda }^{y} ,d_{\lambda }^{z} ),} \end{array}
\end{equation}
where $\vec{d}_{f,\lambda } $ are the model parameters, \textbf{ $\lambda =\pm ,0$.}

In this approach the averaged over the configurations effective dipole moment of a tetrahedron $\left\langle \vec{P}^{B} \right\rangle _{c} $
and the Edwards-Anderson parameter $Q_{EA} =Q_{EA,f} $ describing the proton disorder on the hydrogen bonds + -- =\dots O,
read ($\eta _{f} =\eta $ for all compositions except the crystals of the AFE region of the  $T-x$ phase diagram)

\begin{equation} \label{GrindEQ_3}
\begin{array}{l} {\left\langle P^{B} \right\rangle _{c} \approx 4\left\langle d_{f}^{z} \right\rangle _{c} \cdot \eta ;\quad \eta =\left\langle th\left(\beta \tilde{\xi }\right)\right\rangle _{\sigma ,g} ;\quad Q_{EA} =\left\langle \left(th\left(\beta \tilde{\xi }\right)\right)^{2} \right\rangle _{\sigma ,g} -\eta ^{2} } \\ {\tilde{\xi }=2\varphi +\varphi _{L} +\left\langle d_{f}^{z} \right\rangle _{c} \left\langle \psi _{6,f} \right\rangle _{c} \varepsilon _{6} +\sigma +\left\langle d_{f}^{z} \right\rangle _{c} \left(1+A_{\psi _{6} } \varepsilon _{6} \right)g;} \\ {\left\langle th^{n} \left(\beta \tilde{\xi }\right)\right\rangle _{\sigma ,g} =\int _{-\infty }^{\infty }d\sigma  \int _{-\infty }^{\infty }dg\; \; R\left(\sigma ,2q+q_{L} \right)\cdot R\left(g,\left\langle G^{2} \right\rangle _{c} \right) \cdot th^{n} \left(\beta \tilde{\xi }\right).\; \quad \quad } \end{array}
\end{equation}
Here $\eta _{f} $ is the average parameter of proton ordering for
the bond $f$; $\varphi =\left\langle \varphi _{f} \right\rangle
_{c} $ is the average value of the cluster field; $\varphi _{L}
=\left\langle \varphi _{L,f} \right\rangle _{c} $ is the average
value of the long-range field. Averaging $\left\langle
F\left(\beta \tilde{\xi }\right)\right\rangle _{\sigma ,g} $ in
(\ref{GrindEQ_3}) are performed over two random fields $\sigma$
and g with Gaussian distribution densities
 $R\left(u,Q\right)=1/\sqrt{2\pi Q} \cdot \exp \left\{-u^{2} / 2Q \right\} $.
One of the weight functions is averaged over the stochastic
deformation fields $g$ ($u\to g$ ) with dispersion $Q\to Q_{g}
=\left\langle G^{2} \right\rangle _{c} \cdot x(1-x)$
($\left\langle G^{2} \right\rangle _{c} $ is the composition and
temperature independent model parameter), another one is averaged
over the stochastic fields $\sigma $ with the dispersion $2q+q_{L}
$.  Here $q=\left\langle \varphi _{f}^{2} \right\rangle _{c}
-\varphi ^{2} $) is the cluster field variance;
  $q_{L} =\left\langle \varphi _{L,f}^{2} \right\rangle _{c} -\varphi _{L}^{2} $ is the long-range field variance.
The parameter $A_{\psi _{6} } $ describes the coupling between fluctuations of the random fields $\psi _{6,f} $ and $G_{f}^{z} $.

The system state is determined by two variational parameters $\varphi$  and \textit{q} -- these
are the parameters, determining  QEA. The expression for the free energy also contains the parameters
$\varphi _{L} $ and $q_{L} $ introduced to take into account the long-range field. These, however, can be expressed via \textit{ $\varphi$  }and \textit{q}
and exclude from consideration.

The parameter of the cluster field fluctuation $q$ is different
from zero at compositions except for the pure systems (x = 0,0 and
x = 1,0). In these compounds $\eta$ and $\varphi$ differ from zero
in the low-temperature phases, whereas in the paraelectric phase
$\eta =0;\; \varphi =\varphi _{L} =0$.

The thermodynamic characteristics of the studied system can be obtained from the thermodynamic Gibbs potential
\begin{equation} \label{GrindEQ__4_}
G(T,\varphi ,q,\varepsilon _{6} )=\frac{1}{2} \bar{c}_{66}^{0} \cdot \varepsilon _{6}^{2} -\bar{e}_{36}^{0} \varepsilon _{6} \cdot E^{z} -\frac{1}{2} \bar{\chi }_{33}^{0} \cdot (E^{z} )^{2} +F_{s} (T,\varphi ,q,\varepsilon _{6} )-\sigma _{6} \cdot \varepsilon _{6}
\end{equation}

Here the three first terms correspond to the average lattice free
energy; $F_{s} (T,\varphi ,q,\varepsilon _{6} )$ is the free
energy of the proton subsystem, calculated with the Hamiltonian
\ref{GrindEQ__1_}, $\bar{c}_{66}^{0} $, $\bar{e}_{36}^{0} $ and
 $\bar{\chi }_{33}^{0} $ are the ``seed'' elastic constant, piezoelectric coefficient, and
 longitudinal susceptibility, respectively; $\sigma _{6} $ is the stress tensor component. Equations for the parameters
 \textit{$\varphi$,} \textit{q }, and  \textit{$\varepsilon _{6} $} are found from the condition of the potential $G(T,\varphi ,q,\varepsilon _{6} )$ extremum.
 Dielectric, piezoelectric, and elastic characteristics of the mixed K$_{1-x}$(NH$_4$)$_x$H$_2$PO$_4$ crystals can be calculated
 using the following relations

\[\chi _{33}^{\varepsilon } =(\frac{\partial P}{\partial E^{z} } )_{\varepsilon _{6} } ; \chi _{33}^{\sigma } =(\frac{\partial P}{\partial E^{z} } )_{\sigma _{6} } ; d_{36} =(\frac{\partial \varepsilon _{6} }{\partial E^{z} } )_{\sigma _{6} } ; c_{66}^{0} =(\frac{\partial \sigma _{6} }{\partial \varepsilon _{6}^{} } )_{E^{z} } ; e_{36} =(\frac{\partial \sigma _{6} }{\partial E^{z} } )_{\varepsilon _{6} } \]
Here $\chi _{33}^{\sigma } $ and $\chi _{33}^{\varepsilon } $ are the susceptibilities of free and clamped crystals
($\varepsilon _{33}^{\sigma } =\bar{\varepsilon }_{33}^{0} +4\pi \chi _{33}^{\sigma } $ and
 $\varepsilon _{33}^{\varepsilon } =\bar{\varepsilon }_{33}^{0} +4\pi \chi _{33}^{\varepsilon } $ are the corresponding permittivities; $d_{36} $ is
 the coefficient of piezoelectric strain; $c_{66}^{E} $ is the elastic constant, and $e_{36}^{} $ is piezoelectric coefficient.

A detailed fitting procedure, providing a good description of all available experimental data, will be performed in our
subsequent paper. Here we shall limit our consideration by a qualitatively analysis of the theory predictions.

\begin{figure}[!h]
\begin{center}
 \includegraphics[scale=0.9]{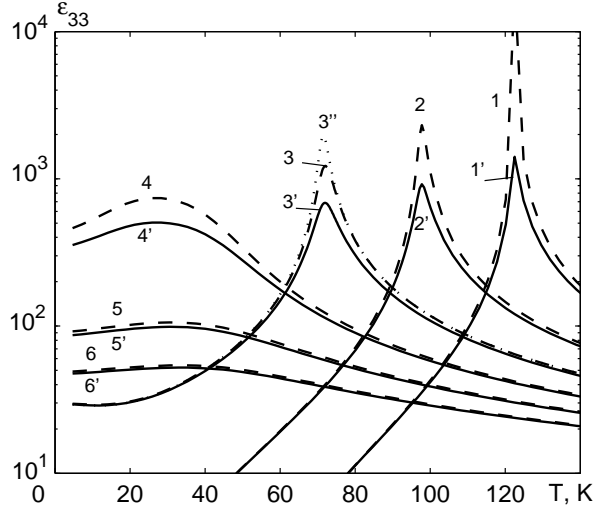}
\end{center}
\caption[]{The temperature dependences of the dielectric permittivities at constant strain $\varepsilon _{33}^{\varepsilon } $ (solid line) and constant stress
 $\varepsilon _{33}^{\sigma } $ (dashed line) for different compositions \textit{x}=0 (1, and 1'); 0.1 (2, and 2');
 0.2 (3, 3' and 3''); 0.3 (4, and 4'), 0.4 (5, and 5'), and 0.5 (6, and 6'), calculated with the variance coefficient $A_{\psi _{6} } =1.$
 The curve 3'' for $x=0.2$ shows the dependence  $\varepsilon _{33}^{\sigma } $(T), calculated for $A_{\psi _{6} } =0.5$ }
 \label{e33_a1_J1_1_J2_04_I_1_Qg_05_x04_1_Pr1_kdp4}
\end{figure}

\begin{figure}[!h]
\begin{center}
 \includegraphics[scale=0.9]{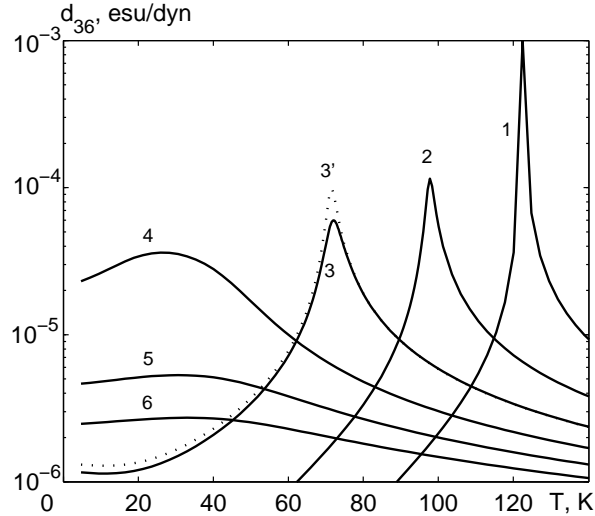}
\end{center}
\caption[]{The temperature dependence of the piezoelectric coefficient $d_{36} $ for different compositions \textit{x}=0 (1); 0.1 (2); 0.2 (3 and 3');
0.3 (4), 0.4 (5), and 0.5 (6),
  calculated with the variance coefficient $A_{\psi _{6} } =1.$ The curve 3'' for $x=0.2$ shows the dependence  $\varepsilon _{33}^{\sigma } $(T), calculated for $A_{\psi _{6} } =0.5$ }
 \label{d36_a1_J1_1_J2_04_I_1_Qg_05_x04_1_Pr1_kdp4}
\end{figure}

\begin{figure}[!h]
\begin{center}
 \includegraphics[scale=0.9]{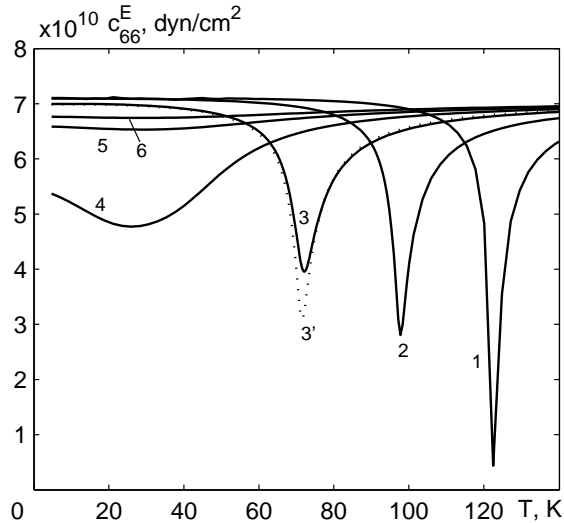}
\end{center}
\caption[]{The temperature dependence of the elastic constant $c_{66}^{E} $  for different compositions ь = 0,0 (1); 0.1 (2); 0.2 (3 and 3'); 0.3 (4),
0.4 (5), and 0.5 (6),
 calculated with the variance coefficient $A_{\psi _{6} } =1$. The curve 3'' for $x=0.2$ shows the dependence  $\varepsilon _{33}^{\sigma } $(T), calculated for $A_{\psi _{6} } =0.5$. }
 \label{c66_a1_J1_1_J2_04_I_1_Qg_05_x04_1_Pr1_kdp4}
\end{figure}

 In figs.~\ref{e33_a1_J1_1_J2_04_I_1_Qg_05_x04_1_Pr1_kdp4} - \ref{c66_a1_J1_1_J2_04_I_1_Qg_05_x04_1_Pr1_kdp4}
 we show that temperature dependences of the dielectric permittivities, piezoelectric coefficient, and elastic constant
 calculated within the proposed model for the K$_{1-x}$(NH$_4$)$_x$H$_2$PO$_4$ crystal.
 One can see that the theoretical and experimental curves are qualitatively similar.


 In particular, the experiment revealed a decrease of the temperature and magnitude of the anomalies in
 $\varepsilon _{33}^{\sigma } $, $d_{36} $, and $c_{36}^{E} $, increase of their smearing with increasing  x,
 which agrees with the theoretical results (figs.~\ref{e33_KDP_Kor_prep} - \ref{c66_KDP_Kor_prep} and figs.~\ref{e33_a1_J1_1_J2_04_I_1_Qg_05_x04_1_Pr1_kdp4}
-\ref{c66_a1_J1_1_J2_04_I_1_Qg_05_x04_1_Pr1_kdp4}). Let us note
that the smearing of the anomalies observed in the phase
transition regions is traditionally ascribed to the spatial
fluctuations of the composition, leading to coexistence of polar
and frustrated paraelectric phases. The proposed theory suggests a
different mechanism of the smearing, which allows us to describe
the temperature curves of $\varepsilon _{33}^{\sigma } $, $d_{36}
$, and $c_{36}^{E} $ without taking into account of the
heterophase fluctuations in the transition region.

Within the proposed model the smearing is attributed to the local fluctuations of the piezoelectric coupling described by the variance
 $A_{\psi _{6} } \cdot Q_{G} $, where $Q_{G} =\left\langle d_{f}^{z} \right\rangle _{c} \sqrt{\left\langle G^{2} \right\rangle _{c} } \cdot x(1-x)$
 is the variance of local deformational fields. The latter is the
 reason why the parameter $Q_{EA} (T,x)$ is different from zero at temperatures well above the transition.
 The calculations show that with increasing the variance coefficient $A_{\psi _{6} } $ the peaks of the dielectric permittivities
 $\varepsilon _{33}^{\sigma } $ and $\varepsilon _{33}^{\varepsilon } $ become more smeared. This is illustrated in
 figs.~\ref{e33_a1_J1_1_J2_04_I_1_Qg_05_x04_1_Pr1_kdp4} - \ref{c66_a1_J1_1_J2_04_I_1_Qg_05_x04_1_Pr1_kdp4},
$d_{36} $ (fig.~\ref{d36_a1_J1_1_J2_04_I_1_Qg_05_x04_1_Pr1_kdp4}), and
 where the temperature curves of $\varepsilon _{33}^{\sigma } $ (fig.~\ref{e33_a1_J1_1_J2_04_I_1_Qg_05_x04_1_Pr1_kdp4}),
and  $c_{36}^{E} $ (fig.~\ref{c66_a1_J1_1_J2_04_I_1_Qg_05_x04_1_Pr1_kdp4})
are shown for the sample with $x = 0.20$ and for the  variance coefficient  $A_{\psi _{6} } = 0,5$ and 1,0.

The performed analysis shows that fluctuations of the piezoelectric coupling smears out the boundary between the ferroelectric and glass-like regions
of the phase diagram, which has been also revealed experimentally (see, for instance, \cite{Korotkov920}).

\section{Conclusions}

In the present paper we describe the thermodynamic approach to description of electromechanical phenomena in ferroelectric crystals.
Relations between the dielectric, piezoelectric, and elastic characteristics of these systems are obtained. The major advantage of the thermodynamic
theory of the ferroelectircs lies in its simplicity, wide application range, and possibility to establish relations between
various macroscopic parameters of the ferroelectrics.  However, it is limited by a purely  macroscopic picture, which makes impossible any
discussion of the microscopic transition mechanisms or of the atomic processes responsible for the ferroelectricity. In fact, this theory is
phenomenological.

The phenomenological description of the physical characteristics of the KH$_2$PO$_4$ type ferroelectrics has been generalized by
taking into account the piezoelectric coupling. However, because of the first order phase transitions in the  KH$_2$PO$_4$  crystals and
large value of the polarization jump at   $T=T_{c}$, the phenomenological approach in this case is quite approximate.

The microscopic approach to description of the thermodynamic and dynamic properties of the ferroelectric crystals of the  KH$_2$PO$_4$
crystals used in this paper is based on the proposed modified proton ordering model with taking into account the linear in the strain $\varepsilon_{6 }$
contributions into the proton subsystem energy; tunneling is neglected. In the four-particle cluster approximation the dielectric, piezoelectric,
elastic and dynamic characteristics of the KH$_2$PO$_4$ type ferroelectrics and NH$_4$H$_2$PO$_4$ type antiferroelectrics have been calculated. Optimum values
of the fitting parameters have been found, providing a proper description of the available experimental data for the considered crystals.

Taking into account the piezoelectric coupling allowed us to calculate the susceptibilities of mechanically free and clamped crystals, piezoelectric
coefficients, elastic constants. The calculated temperature dependences of these characteristics confirmed the experimentally observed
difference between $\varepsilon_{33}^{\sigma}$ and $\varepsilon_{33}^{\varepsilon}$ for the K(H$_{1-x}$D$_{x}$)$_2$PO$_4$ type crystals and
a small difference between them for ADP and DADP.

 It has been shown that in presence of the piezoelectric coupling the minimal value of $\varepsilon'_{33}(\omega)$ at different frequencies is larger
 than in its absence, in agreement with experiment. With increasing $\varepsilon_{6}$ the dispersion frequency of $\varepsilon_{33}(\omega)$ in ferroelectric
 crystals of the KH$_2$PO$_4$ family increases as well. Phenomena of crystal clamping and piezoelectric resonance in the crystals have been described for the
 first time.

Using the thermodynamic theory and the obtained experimental data for the resonance $f_r$  and antiresonance $f_a$ frequencies, the coefficients of
piezoelectric coupling, longitudinal dielectric permittivities of free and clamped crystals, coefficient of piezoelectric strain $d_{36}$, elastic
constant at constant field $c_{66}^{E}$ have been calculated. A typical for the piezoelectrics behavior of these characteristics in the
 K$_{1-x}$(NH$_4$)$_x$H$_2$PO$_4$ systems has been revealed.

It is shown that with increasing ammonium content for the studied crystals, the anomalies of $\varepsilon _{33}^{\sigma } $, $d_{36} $, and $c_{66}^{E} $
in the phase transition regions  are getting more and more smeared out. It is establised that these disordered compounds are piezoelectric.

In order to describe the dielectric, piezoelectric, and elastic
characteristics of the mixed crystals of the
K$_{1-x}$(NH$_4$)$_x$H$_2$PO$_4$  type we propose a model, which
takes into account the piezoelectric coupling, in addition to the
competing long-range and short-range interactions and random
internal field. The dielectric permittivities of free and clamped
crystals, coefficient of piezoelectric strain  $d_{36} $, and the
elastic constant $c_{66}^{E}$ are calculated and explored. It is
shown that the theoretical results are in a qualitative agreement
with experimental data.

The analysis performed within the framework of the proposed model revealed a possible additional origin for the observed smearing of the  anomalies in the
temperature dependences of the dielectric, piezoelectric, and elastic characteristics in the vicinities of the transitions into the ferroelectric or
glass-like phases, which is related to the spatial dispersion of the coefficient of piezoelectric coupling.

Without a doubt,  further theoretical and experimental studies of
the thermodynamic, dielectric, piezoelectric, and elastic
characteristics of the Rb$_x$(NH$_{4})_{1-x}$H$_2$PO$_4$ and
K$_x$(NH$_{4})_{1-x}$H$_2$AsO$_4$ systems. Particularly useful
information about the
 K$_x$(NH$_{4})_{1-x}$H$_2$AsO$_4$ type systems can be obtained from the studies of the transverse dielectric, piezoelectric, and elastic characteristics.

\section*{Acknowledgement}
The authors acknowledge financial support from the Russian Foundation for Basic Research, Project No. 13-02-90473/A and from the
State Foundation for Fundamental Studies of Ukraine, Project No. F53.2/070.


\begin{thebibliography}{99}

\bibitem{Vdovych111} Levitskii R.R., Sorokov S.I., Vdovych A.S. Spin model with
different types of competing interactions. // Ferroelectrics. --
2005. -- Vol. 316. -- P. 111-119.

\bibitem{Vdovych603} Sorokov S.I., Levitskii R.R., Vdovych A.S. Spin-glass model
with essential short-range competing interactions. // Condens.
Matter Phys. -- 2005, -- Vol. 8, No. 3(43). -- P. 603-622.

\bibitem{VdovychJPS} Sorokov S.I., Vdovych A.S., Levitskii R.R. // J. Phys. Studies. - 2009.
- vol.13,  p. 1701.

\bibitem{Vdovych523} Levitskii R.R., Sorokov  S.I., Stankowski J.,
Trybula Z., Vdovych A.S. Termodynamics and complex dielectric
permittivity of mixed crystals of the
Rb$_{1-x}$(NH$_{4}$)$_{x}$H$_{2}$PO$_{4}$ type. // Condens. Matter
Phys. -- 2008. -- Vol. 11, No. 3(55). -- P. 523-542.

\bibitem{cmp13706}  Sorokov S.I., Levitsky R.R., Vdovych A.S. Thermodynamics, dielectric permittivity and phase diagrams of the  Rb$_{1-x}$(NH$_{4}$)$_{x}$H$_{2}$PO$_{4}$ type proton glasses // Condens. Matter Phys. - 2010. - Vol. 13, No 1. - P. 13706: 1-26.

\bibitem{Vdovych101} Sorokov S.I., Levitskii R.R., Vdovych A.S. Microscopic Theory of
Rb$_{1-x}$(NH$_{4}$)$_{x}$H$_{2}$PO$_{4}$ Type Compounds //
Ferroelectrics. -- 2009. -- Vol. 379, Issue 1. -- P. 101 - 106.

\bibitem{ferro43}   Sorokov S. I., Levitsky R.R., Vdovych A.S., Korotkov L.N. Thermodynamic and Dielectric Properties of K$_{1-x}$(NH$_{4}$)$_{x}$H$_{2}$PO$_{4}$  Mixed Crystal // Ferroelectrics. - 2010. - Vol. 397, Issue 1. - P. 43 - 53.

\bibitem{izv1268} Sorokov S.I. // Bull. RAS, ser. phys.,- 2010 - vol.74, -p.1268-1272.

\bibitem{icmp1113e} Levitskii R.R., Sorokov S.I.,  Zachek I.R., Vdovych A.S., Moina A.P., Korotkov L.M., Bocharov A.I.  - Lviv, 2011. - 112p.  (Prepr. / NAN of Ukraine. Institute for Condensed Matter Physics; ICMP-11-13E).

\bibitem{Tu93} Tu C.S., Schmidt V.H., Saleh A.A. Dielectric relaxation and
piezoelectric coupling in the mixed proton-glass crystall
K$_{0.61}$(NH$_{4})_{0.39}$H$_{2}$PO$_{4}$ // Phys. Rev. B. - 1993.
-- Vol. 48, No 17. -- P. 12483-12487.

\bibitem{He91} He P., Deguchi K., Nakamura E.  // J. Phys. Soc. Jpn. --
1991. -- Vol. 60, No. 7. -- P. 2143-2146.



\bibitem{16x} V. G. Vaks, Introduction into Microscopic Theory of Ferroelectrics
~Nauka, Moscow, 1973 (in Russian).

\bibitem{133x} Levitskii R.R., Korinevskii N.A., Stasyuk I.V.
// Ukr. Journ. Physics. - 1974. -vol.19,  - p.1289-1297.

\bibitem{17x} R. Blinc and B. Zeks, Soft Modes in Ferroelectrics and Antiferroelectrics
~Elsevier, New York, 1974.

\bibitem{3lbm} Levitskii R.R., Lisnii B.M., Baran O.R. Thermodynamics and
dielektric properties of KH$_{2}$PO$_{4}$, RbH$_{2}$PO$_{4}$,
KH$_{2}$As0$_{4}$, RbH$_{2}$As0$_{4}$ ferroelectrics // Condens.
Matter Phys. - 2001. - Vol. 4, - P. 523-552.


\bibitem{471x} Levitskii R.R., Lisnii B.M. // Journ. Phys. Studies. - 2002. - vol. 6,
- p. 91-108.


\bibitem{5lbm} Levitskii R.R., Lisnii B.M., Baran O.R. Thermodynamics and
dielektric properties of the NH$_{4}$H$_{2}$PO$_{4}$ type
antiferroelectrics // Condens. Matter Phys. - 2002. - Vol. 5, No
3. - P. 553-577.

\bibitem{uni2009} Levitskii R.R., Zachek I.R., Vdovych A.S., Sorokov S.I. Thermodynamics and dynamical properties of the KH$_2$PO$_4$ type
ferroelectric compounds. A unified model // Condens. Matter Phys. --
2009. -- Vol. 12, No 1, pp. 75-119.


\bibitem{0311U1}
Yomosa Sh., Nagamiya T. The phase transition and the piezoelectric
effect of KH$_2$PO$_4$. // Progr. Theor. Phys., 1949, v.~4, No~3,
p.~263--274.

\bibitem{0311U2}
Slater J.C. Theory of the transition in KH$_2$PO$_4$.
// J. Chem. Phys., 1941, v.~9, No~1, p.~16--33.


\bibitem{Izv705}
I.V. Stasyuk and I.N. Biletskii, Bull. Acad. Sci. USSR, Phys. Ser.
\textbf{4}, 79 (1983).

\bibitem{0311U5}
Stasyuk I.V., Biletskii I.N., Styahar O.N.
// Ukr. Journ. Phys., 1986, vol.~31,  p.~567--571.

\bibitem{0311U6}
Stasyuk I.V., Levitskii R.R., Zachek I.R., Moina A.P. The
KD$_2$PO$_4$ ferroelectrics in external fields conjugate to the
order parameter: Shear stress $\sigma_6$.
// Phys. Rev. B, 2000, v.~62, No.~10, p.~6198--6207.

\bibitem{JPS1701}   Levitsky R.R., Zachek I.R., Vdovych A.S., Moina A.P.
Longitudinal dielectric, piezoelectric, elastic, and thermal characteristics of the KH$_2$PO$_4$ type ferroelectrics
// J. Phys. Study. - 2010. - Vol. 14, No 1. - P. 1701.

 \bibitem{Lis2003}
 Levitskii R.R., Lisnii B.M., // Journ. Phys. Studies, 2003, vol.~7,
 p.~431-445.

 \bibitem{Lis2004PSS}
Levitskii R.R., Lisnii B.M. Theory of related to shear strain
$u_6$ physical
 properties of ferroelectrics and antiferroelectrics of the KH$_2$PO$_{4}$
family // phys. stat. sol. (b). - 2004. -Vol.241, No
6.-P.1350-1368.

\bibitem{Lisnyj2004}
 Lisnii~B.M., Levitskii~R.R. Theory of physical properties of
 ferro- and antiferroelectrics of the KH$_2$PO$_4$ family related
 to strains $u_4$ and $u_5$ // Ukr. J. Phys., 2004, v.~49, No 7,
 p.701-709.

\bibitem{0311U7}
Stasyuk I.V., Levitskii R.R., Moina A.P., Lisnii B.M. Longitudinal
field influence on phase transition and physical properties of the
KH$_2$PO$_4$ family ferroelectrics.
// Ferroelectrics, 2001, v.~254, p.~213--227.

\bibitem{PCSS635} Levitskii R.R., Zachek I.R., Vdovych A.S. // Physics and Chemistry of Solid State. - 2009. - vol. 13, p. 635-646.

\bibitem{Lis2007}
Lisnii~B.M., Levitskii~R.R., Baran O.R. // Phase Transitions. 2007.  \textbf{80}. N 1-2.
25-30.

\bibitem{UFJ2008} Stasyuk I.V., Levitskii R.R., Moina A.P., Velychko O.V. //
Ukr. Journ. Phys.: Reviews. 2008. No 1. p.3-6.



\bibitem{cmp555} Levitsky R.R., Zachek I.R., Moina A.P., Vdovych A.S. Longitudinal relaxation of mechanically free KH$_2$PO$_4$ type crystals.
Piezoelectric resonance and sound attenuation // Condens. Matter Phys. - 2008. - Vol. 11, No 3(55). - P. 555-570.

\bibitem{PCSS377}   Levitskii R.R., Zachek I.R., Vdovych A.S. // Physics and Chemistry of Solid State. - 2009. - vol. 10, p. 377-388.

\bibitem{cmp275} Levitsky R.R., Zachek I.R., Moina A.P., Vdovych A.S. Longitudinal relaxation of ND$_4$D$_2$PO$_4$ type antiferroelectrics. Piezoelectric resonance and sound attenuation // Condens. Matter Phys. - 2009. - Vol. 12, No 2. - P. 275-294.

\bibitem{cmp33705} Levitskii R.R., Zachek I.R., Vdovych A.S. Longitudinal relaxation of mechanically clamped KH$_2$PO$_4$ type crystals // Condens. Matter Phys. - 2012. - Vol. 15, No 3. - P. 33705: 1-20.




\bibitem{483x} Ono Y., Hikita T., Ikeda T.  // J. Phys. Soc. Jpn. --
1987. -- Vol. 56, No. 2. -- P. 577-588.

\bibitem{Gridnev91} Gridnev S.A., Korotkov L.N., Shuvalov L.A., Rogova
S.P.,Fedosyuk R.M.
// Ferroelectrics Lett. -- 1991. -- Vol. 13. -- P. 67-72.

\bibitem{Korotkov2004} Korotkov L. N. and Shuvalov L. A.   // Crystallography Reports. 2004. Vol. 49. P. 832

\bibitem{Lines1977} L.E. Lines and A.M. Glass, Principles and Application of Ferroelectrics and Related Materials, Oxford: Clarendon (1977).




%\bibitem{Tu93}
%C.-S.Tu, V.H. Schmidt, A.A. Saleh, Phys. Rev. B, 1993,
%\textbf{48}, 12483.
%
%\bibitem{He91} He P., Deguchi K., Nakamura E.  // J. Phys. Soc. Jpn. --
%1991. -- Vol. 60, No. 7. -- P. 2143-2146.

\bibitem{NULP127} Levitskii R.R., Vdovych A.S., Zachek I.R. //
J. of National Univ. LLvivska PolitechnikaL. - 2011. - vol. 696,  - p. 127 - 135.

\bibitem{izv1414}   Levitskii R.R., Zachek I.R.,Korotkov L.N.,Vdovych A.S., Sorokov S.I.   // Izvestiya RAN. Seriya Fizicheskaya. - 2011. - Vol.  75, N 10. - pp. 1473-1478.



\bibitem{Korotkov1120} Korotkov L.N., Likhovaya D., Sorokov S.I., Levitskii R.R.,
Vdovych A.S., Trybula Z., Los Sz., Zakhvalinskii V.S., Khmara A.N., Pilyuk E.A., Sitalo E.I.
// Bull. RAS, ser. phys. - 2013, vol. 77, p. 1120-1125.

\bibitem{Korotkov52}  Korotkov L.N., Likhovaya D.V., Levitskii R.R., Sorokov S.I., Vdovych A.S. Anomalies of dielectric, elastic and electromechanical properties of K$_{0.25}$(NH$_4$)$_0.75$H$_2$PO$_4$  single crystal in the vicinity of antiferroelectric phase transition // Solid State Commun. - 2013. - Vol.160. - P. 52-55.

\bibitem{Korotkov76}  Korotkov L., Likhovaya D., Levitskii R., Sorokov S., Vdovych A. Dielectric, Elastic and Electromechanical Properties of K$_{1-x}$(NH$_4$)$_x$H$_2$PO$_4$ Solid Solutions in Paraelectric Phase // Ferroelectrics. - 2013. - Vol.444. - P.76-83.









\bibitem{Matsushita85} Matsushita E., Matsubara T.  // J.Phys.Soc.Jap.
-- 1985. -- Vol. 54, N 3. -- P. 1161-1167.

\bibitem{Pirc87} Pirc R., Tadic B., and Blinc R.  // Phys.Rev.B. -- 1987. -- Vol. 36, N 16.
-- P. 8607-8615.

\bibitem{Banerjee2003} Banerjee V., Dattagupta S.  // Phys.Rev B. -- 2003. -- Vol. 68. -- P. 054202.









\bibitem{143x} I.S. Zheludev, Physics of crystalline dielectrics (Plenum Press, 1971).


\bibitem{719x} W. G. Cady, Piezoelectricity; An Introduction to the Theory and
Application of Electromechanical Phenomena in Crystals
(McGraw Hill, New York, 1946).

\bibitem{138x} W. P. Mason, Piezoelectric Constants and Their Application to
Ultrasonics (Van Nostrand, New York, 1950).

\bibitem{139x}   W.Kanzig,
Ferroelectrics and Antiferroelectrics (Academic Press, New York, 1957).

\bibitem{142x} F.Jona and G. Shirane, Ferroelectric Crystals (Pergamon Press,
Oxford, 1962).

\bibitem{352x}  J. Burfoot. Ferroelectrics. An introduction to the physical Principles. (Van Nostrand, Princeton,  1967).

\bibitem{147x} M. E. Lines and A. M. Glass, Principles and Application of Ferroelectrics and Related Materials (Clarendon
Press, Oxford, 2001).

\bibitem{151x}
G. A. Smolenskii, V. A. Bokov, V. A. Isupov, N. N. Kraini, P. E. Pasynkov and A. I. Sokolov:
Ferroelectrics and Related Materials, ed. G. Taylor and G. A. Smolenskii (Gordon and Breach, New York, London, Paris, Montreux, Tokyo, 1984).









\bibitem{cond2012} R.R.Levitskii, I.R.Zachek, A.P. Moina, A.S. Vdovych
Piezoelectric Resonance in KH2PO4  Type Crystals Revisited ar.xiv:1212,4506 v1 [cond-mat.mtrl-sci] 18 Dec 2012, P.1-5

\bibitem{0608U} R.R. Levitskii, I.R. Zachek, A.S. Vdovych. Preprint ICMP-06-08U, Lviv, 2006.  117 p.


\bibitem{0819U} R.R. Levitskii, I.R. Zachek, A.S. Vdovych. Preprint ICMP-08-19U, Lviv. - 61 p.



\bibitem{shuv66} Shuvalov~L.A., Mnatsakanyan~A.V.  // Sov. Phys. Crystall., 1966, vol.
11, No 2, p. 210-212.




\bibitem{371x} Shuvalov L.A., Zheludev I.S., Mnatsakanyan A.V., Ludupov Ts. Zh.,
Fiala I.
// Bull. Ac. Sci. USSR, ser. phys. - 1967. - vol.31, p.1919-1922.


\bibitem{367x} Strukov B.A., Baddur A., Velichko I.A. // FTT - 1971.-
vol .13, No 8. - p. 2484-2485.

\bibitem{403x} Pereverzeva L.P. //
Bull. Ac. Sci. USSR, ser. phys. - 1971. - vol.35, No 12. - T. 2613-2614.

\bibitem{379x} Vasilevskaya A.S., Sonin A.S.
// FTT. - 1971. -vol.13, No 6. - p. 1550-1556.




\bibitem{48pok} \textit{Stasyuk I.V., Levitskii R.R., Korinevskii N.A.}  Phys. Stat. Sol. (b).  1979. \textbf{91}. N 2.  541-550.

\bibitem{135x} \textit{Levitskii R.R., Stasyuk I.V., Korinevsky H.A.}  Ferroelectrics.
1978.  \textbf{21}.  481-483.

\bibitem{137x} \textit{Korinevskii N.A., Levitskii R.R.}  Theoret. Math. Phys..  1980.
\textbf{42}. p. 416-429.

\bibitem{485x} Matthias T., Merz W., Scherrer P.
// Helv. Phys. Acta. - 1947. - Vol. 20. - P. 273-306.

\bibitem{bantle43} \textit{Bantle W., Caflish C.}  Helv. Phys. Acta. 1943. \textbf{16}.  235.

\bibitem{von43} \textit{Von Arx A., Bantle W.}  Helv. Phys. Acta. 1943. \textbf{16}. 211.

\bibitem{brody68} \textit{Brody E.M., Cummins H.Z.} Phys. Rev. Lett. 1968. \textbf{21}. 1263.

\bibitem{385x} \textit{Garland C.W., Novotny D.B.}  Phys.Rev. 1969. \textbf{177}. N 2.  971-975.

\bibitem{398x} \textit{Gauss K.E., Happ H., Rother G.}  Phys.
Stat. Sol.~B.  1975.  \textbf{72}. N 2.  623-630.

 \bibitem{408x}
 \textit{Volkov A.A., Kozlov G.V., Lebedev S.P., Velichko I.A.}
 FTT, 1979. \textbf{21}. p. 3304-3309.

\bibitem{465x} \textit{Kozlov G.V., Lebedev S.P., Prokhorov A.M., Volkov A.A.}
 J.Phys.Soc.Japan.  1980. \textbf{49}.  Suppl. 188-190.



\bibitem{Korotkov920} Korotkov L.N., Shuvalov L.A. // Crystallography, 2004. vol. 49. p. 920.













\end{thebibliography}
\end{document}